\pdfoutput=1

\documentclass[12pt]{article}
\usepackage{amsfonts}
\usepackage{amsmath}
\usepackage{amssymb}
\usepackage{bigints}
\usepackage{booktabs}
\usepackage[nosort]{cite}
\usepackage{color}
\usepackage{dsfont}
\usepackage{float}
\usepackage{framed}
\usepackage{graphicx}
\usepackage{indentfirst}
\usepackage{mathrsfs}
\usepackage{multirow}
\usepackage{pdflscape}
\usepackage{setspace}
\usepackage{subdepth}
\usepackage{subfig}
\usepackage{titlesec}
\usepackage[dotinlabels]{titletoc}
\usepackage{wrapfig}
\usepackage[all]{xy}
\usepackage{young}
\usepackage[vcentermath]{youngtab}
\usepackage{relsize}
\usepackage{stackengine}
\usepackage{datetime}

\usepackage{hyperref}

\usepackage{bbm}

\numberwithin{equation}{section}

\usepackage{verbatim}

\newcommand{\be}{\begin{equation}}
\newcommand{\ee}{\end{equation}}

\newcommand{\bml}{\begin{multline}}
\newcommand{\emll}{\end{multline}}

\newcommand{\nn}{\nonumber}

\def\({\left(} \def\){\right)}
\def\[{\left[} \def\]{\right]}

\def\mO{\mathcal{O}}

\def\mM{\mathcal{M}}

\def\lam{\lambda}

\def\D{\Delta}

\def\kk{\text{,}}

\newmuskip\pFqmuskip
\newcommand*\pFq[6][8]{%
  \begingroup 
  \pFqmuskip=#1mu\relax
  \mathcode`\,=\string"8000
  \begingroup\lccode`\~=`\,
  \lowercase{\endgroup\let~}\pFqcomma
  {}_{#2}F_{#3}{\left[\genfrac..{0pt}{}{#4}{#5};#6\right]}%
  \endgroup
}
\newcommand{\pFqcomma}{\mskip\pFqmuskip}

\newmuskip\LpFqmuskip
\newcommand*\LpFq[6][8]{%
  \begingroup 
  \pFqmuskip=#1mu\relax
  \mathcode`\,=\string"8000
  \begingroup\lccode`\~=`\,
  \lowercase{\endgroup\let~}\pFqcomma
  {}_{}F_{}{\left[\genfrac..{0pt}{}{#4}{#5};#6\right]}%
  \endgroup
}

\newmuskip\Ftmuskip
\newcommand*\Ft[6][8]{%
  \begingroup 
  \pFqmuskip=#1mu\relax
  \mathcode`\,=\string"8000
  \begingroup\lccode`\~=`\,
  \lowercase{\endgroup\let~}\pFqcomma
  F_{2}{\left[\genfrac..{0pt}{}{#4}{#5};#6\right]}%
  \endgroup
}
\newmuskip\Ftmuskip
\newcommand*\FK[6][8]{%
  \begingroup 
  \pFqmuskip=#1mu\relax
  \mathcode`\,=\string"8000
  \begingroup\lccode`\~=`\, 
  \lowercase{\endgroup\let~}\pFqcomma
  F_{K}{\left[\genfrac..{0pt}{}{#4}{#5};#6\right]}%
  \endgroup
}

\newcommand{\G}{\Gamma}

\newcommand{\la}{\langle}
\newcommand{\ra}{\rangle}

\newcommand{\bea}{\begin{eqnarray}}
\newcommand{\eea}{\end{eqnarray}}

\usepackage[left=2cm,right=2cm,top=2cm,bottom=2cm]{geometry}
\titleformat{\section}{\normalfont\bfseries}{\thesection.}{4pt}{}
\titlespacing{\section}{0pt}{22pt}{6pt}

\titleformat{\subsection}{\normalfont\itshape}{\thesubsection.}{4pt}{}
\titlespacing{\subsection}{0pt}{18pt}{6pt}

\titleformat{\subsubsection}{\normalfont\itshape}{\thesubsubsection.}{4pt}{}
\titlespacing{\subsubsection}{0pt}{16pt}{6pt}

\def\ie{\begin{equation}\begin{aligned}}
\def\fe{\end{aligned}\end{equation}}

\def\tilde{\widetilde}
\def\t{\tilde}

\def\h{\mathbbm}
\def\bar{\overline}

\def\1{{\mathds 1}}

\def\mL{\mathcal{L}}

\DeclareFontShape{OT1}{cmr}{mx}{n}%
    {<->cmr10}{}
\newcommand{\mytitlefont}{\fontseries{mx}\selectfont}
\DeclareMathAlphabet{\titlemath}{OT1}{cmr}{mx}{n}

\begin{document}


\vfill

\begin{titlepage}

\begin{center}

~\\[2cm]

{\fontsize{20pt}{0pt} \mytitlefont Multipoint Conformal Blocks in the Comb Channel}

~\\[0.5cm]

{\fontsize{14pt}{0pt} Vladimir Rosenhaus}

~\\[0.1cm]

\it{School of Natural Sciences, Institute for Advanced Study}\\ \it{1  Einstein Drive, Princeton, NJ 08540}

~\\[0.8cm]

\end{center}

\noindent 

Conformal blocks are the building blocks for correlation functions in conformal field theories. The four-point function is the most well-studied case. We consider conformal blocks for $n$-point correlation functions. For conformal field theories in dimensions $d=1$ and $d=2$, we use the shadow formalism to compute $n$-point conformal blocks, for arbitrary $n$, in a particular channel which we refer to as the comb channel. The result is expressed in terms of a multivariable hypergeometric function, for which we give series, differential, and integral representations. In general dimension $d$ we derive  the $5$-point conformal block, for external and exchanged scalar operators. 
\vfill

\end{titlepage}

\tableofcontents


\section{Introduction}

\begin{figure}[t]
\centering
\subfloat[]{
\includegraphics[width=1.8in]{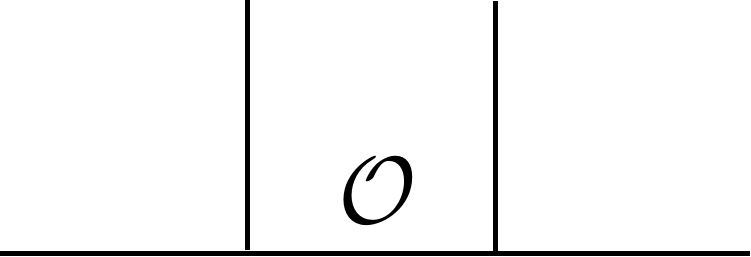}} \ \ \ \ \  \ \ \ \ \ \ \ 
\subfloat[]{
\includegraphics[width=3in]{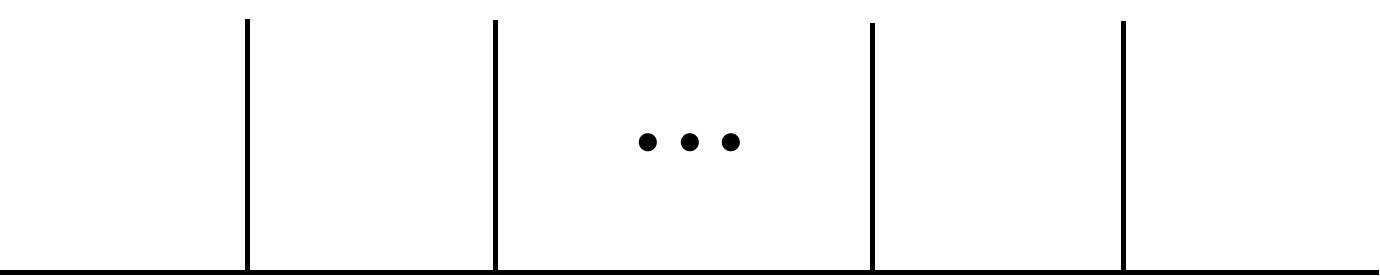}}
\caption{(a) $4$-point conformal block. (b) $n$-point conformal block, in the comb channel. 
}
\end{figure} \label{4ptI}

Correlation functions of local operators are fundamental observables in  quantum field theory. In computing and specifying  correlation functions,  it is useful to exploit the symmetries of the theory to the fullest extent possible. In particular, one would like to write the correlation functions in a way that separates the theory-dependent data from the universal pieces. 

In a $d$ dimensional quantum field theory endowed with conformal symmetry, a CFT,   the theory-dependent data are the dimensions and OPE coefficients of the primary operators, and the universal pieces are the conformal blocks. The conformal blocks provide a bridge between the observables: correlation functions that one measures, either experimentally or theoretically, and the CFT data: dimensions and OPE coefficients of operators. More technically, conformal symmetry fully fixes the OPE coefficients of the descendants in terms of the OPE coefficients of the primaries, and the conformal blocks sum all the descendants.

We can view a four-point conformal block as arising from  taking a four-point function and adding a projector,  in between the second and third operator, onto a particular intermediate operator $\mO$ and its descendants, as shown in Fig.~\ref{4ptI}(a). 
An $n$-point conformal block, such as the one shown in Fig.~\ref{4ptI}(b),  has $n{-}3$ intermediate operators.

Four-point blocks in dimensions $d=1$ and $d=2$ are simple and have been known since the 70's \cite{Ferrara:1974ny, Osborn:2012vt}. The modern study of four-point blocks in higher dimensions was initiated by Dolan and Osborn \cite{Dolan:2000ut, Dolan:2003hv, Dolan:2011dv}. 
 The goal of this work is to take initial steps towards the study of $n$-point conformal blocks, for any $n$. For fixed $n\geq  6$,  there are in fact multiple $n$-point blocks. 
 What is perhaps the simplest channel is shown in Fig.~\ref{4ptI}(b), and we  will refer to it as the comb channel.
 
  In Sec.~\ref{2d} we compute the $n$-point conformal blocks in the comb channel, in dimensions $d=1$ and $d=2$. The result for $d=1$, see Eq.~\ref{nptg0}, is expressed in terms of an $n{-}3$ variable generalized hypergeometric function, which we call the comb function. In Appendix~\ref{appFK} we derive some properties of the comb function. 
 The answer for the blocks, while seemingly complicated,  actually takes the simplest form one could have hoped for, and reflects the symmetry of the comb channel. In a technical sense, the statement of simplicity is that the conformal block is expressed as an $n{-}3$ fold sum, the smallest possible number, as there are $n{-}3$ independent cross-ratios. 
The $n$-point blocks for $d=2$ immediately follow from those for $d=1$, see Eq.~\ref{G2d}. 

In higher dimensions, one expects the blocks to be significantly more involved.  In Sec.~\ref{dd} we begin the study of higher-point blocks in $d$ dimensions, focusing on the simplest case: the $5$-point block with external and exchanged scalar operators. The relevant technical details are in Appendices~\ref{MBapp} and \ref{5dDetails}. We end in Sec.~\ref{dis} with comments on future directions.

\section{One and Two Dimensions} \label{2d}

In this section we compute $n$-point blocks in the comb channel, for any $n$, in one and two dimensions. 
In Sec.~\ref{summary} we  establish notation and summarize the result.  In Sec.~\ref{4ptSec} we review the $4$-point block. In Sec.~\ref{5ptSec} we derive the $5$-point block. In Sec.~\ref{nptSec} we derive the $n$-point blocks in the comb channel. In Sec.~\ref{Analysis} we check that the blocks have the correct  leading OPE behavior and that they satisfy the appropriate Casimir differential equations. 

\subsection{Summary} \label{summary}

\subsubsection{Definitions}
We first discuss conformal blocks in one dimension. An $n$-point conformal block with external operators of dimensions $h_i$ and exchanged operators of dimensions $\h h_i$ will be denoted by $G_{\h h_1, \ldots, \h h_{n-3}}^{h_1, \ldots, h_n}(z_1, \ldots, z_n)$. A $n$-point block can be decomposed into a function of the conformally invariant cross ratios $\chi_i$, of which there are $n{-}3$, and what we will refer to as the leg factor,
\be \label{GDEF}
\!\!\! G_{\h h_1, \ldots, \h h_{n-3}}^{h_1, \ldots, h_n}(z_1, \ldots, z_n)  =\mL^{h_1, \ldots, h_n}(z_1, \ldots, z_n)\, \, g_{\h h_1, \ldots, \h h_{n-3}}^{h_1, \ldots, h_n}(\chi_1, \ldots, \chi_{n-3})~, \ \ \ \ \ \ \ 
\chi_i = \frac{z_{i, i+1}\, z_{i+2, i+3}}{z_{i, i+2}\, z_{i+1, i+3}}~,
\ee
where our notation is  $z_{i j}\equiv z_{i, j} \equiv z_i - z_j$. We will sometimes refer to $ g_{\h h_1, \ldots, \h h_{n-3}}^{h_1, \ldots, h_n}(\chi_1, \ldots, \chi_{n-3})$ as the bare conformal block. The leg factor only depends on the external dimensions. We take it to be,
\be \label{Leg}
\mL^{h_1, \ldots, h_n}(z_1, \ldots, z_n) = \(\frac{z_{23}}{z_{12} z_{13}}\)^{h_1}\(\frac{z_{n-2, n-1}}{z_{n-2, n} \, z_{n-1,n}}\)^{h_n} \, \prod_{i=1}^{n-2} \(\frac{z_{i, i+2}}{z_{i, i+1}\, z_{i+1, i+2}}\)^{h_{i+1}}\,~.
\ee
The leg factor transforms as an $n$-point CFT correlation function and, in particular, has dimension $\sum_{i=1}^n h_i$. The leg factor can of course be changed, by a function of the cross-ratios, at the expense of changing the bare conformal block. The above leg factor will emerge naturally in the derivation of the blocks, and leads to the bare blocks having a simple form.

The conformal Casimir in two dimensions (discussed later in Sec.~\ref{CasimirSec})  factorizes into a sum of two SL$_2$ Casimirs, $l^2 + \bar{l}^2$, built out of $z$ and $\bar{z}$ respectively, and the eigenvalues are a sum, $
\h h( \h h - 1) + \bar{\h h}(\bar{\h h} - 1)$. 
 As a result,  the two-dimensional conformal blocks are simply a product of two one-dimensional conformal blocks, one for the holomorphic sector and one for the antiholomorphic sector,~\footnote{Sometimes it is natural to combine conformal blocks. For instance, for the 4-point block, if all external operators are scalars, $h_i = \bar{h}_i$, then one takes the sum of the block with exchanged operator $(\h h_1, \bar{\h h}_1)$ and the block with exchanged operator $(\bar{\h h}_1, \h h_1)$, and refers to this as the conformal block. This is convenient because then the block has $z\leftrightarrow \bar{z}$ symmetry.}
 \be \label{G2d}
 G_{\h h_1,\ldots \h h_{n-3}; \bar{\h h_1}, \ldots, \bar{\h h}_{n-3}}^{ h_1, \ldots,  h_n; \bar{h}_1, \ldots, \bar{h}_{n}}(z_1, \ldots, z_n; \bar{z}_1, \ldots, \bar{z}_n) = G_{\h h_1, \ldots, \h h_{n-3}}^{h_1, \ldots, h_n}(z_1, \ldots, z_n) \, G_{\bar{\h h}_1, \ldots, \bar{\h h}_{n-3}}^{\bar{h}_1, \ldots, \bar{h}_n}(\bar{z}_1, \ldots, \bar{z}_n)~.
 \ee
 The dimension and spin are defined to be $\D = h + \bar{h}$ and $J = h - \bar{h}$. Since the two-dimensional blocks follow immediately from the one-dimensional blocks,  we can continue to work in one dimension for the rest of this section.~\footnote{In two dimensions, conformal symmetry is enhanced to Virasoro symmetry, giving rise to the study of Virasoro blocks \cite{Belavin:1984vu, Zamolodchikov:1985ie,1987TMP....73.1088Z }. In the limit of large central charge, these reduce to the global conformal blocks, see for instance \cite{Cho:2017oxl, Alkalaev:2015fbw, Fitzpatrick:2015zha}. In the $2d$ language our   blocks are the global conformal blocks. The $5$-point global blocks were computed in \cite{Alkalaev:2015fbw}; our answer  will be simpler. 
}

\begin{figure}
\centering
\includegraphics[width=4in]{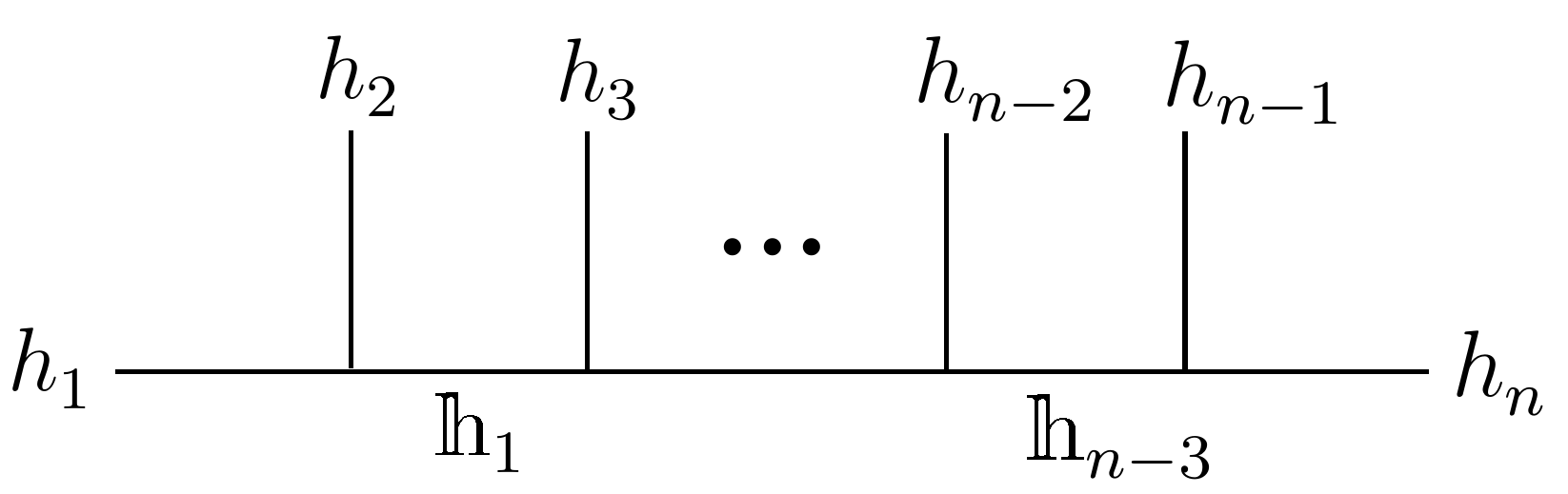}
\caption{The $n$-point conformal partial wave in the comb channel. Each vertex denotes a conformal three-point function. Each line denotes a position. Internal lines denote positions that are integrated over. The $h_i$ denote the  dimensions of the external operators, and the $\h h_i$ are the  dimensions of the exchanged operators. 
} \label{nptF}
\end{figure}

\subsubsection{Conformal partial wave} \label{CPWSec}
The conformal partial wave, to be defined below, is denoted by $\Psi_{\h h_1, \ldots, \h h_{n-3}}^{h_1, \ldots, h_{n}}(z_1, \ldots, z_n)$. The $4$-point partial wave is defined by the following integral,~\footnote{In some places in the literature, what we call a conformal block is called a conformal partial wave, and what we call a bare conformal block is called a conformal block, and (\ref{4ptCPW}) has no name. Our usage follows the standard in recent literature, which is to refer to (\ref{4ptCPW}) as the conformal partial wave.}
\be \label{4ptCPW}
\Psi_{\h h_1}^{h_1, h_2, h_3, h_4}(z_1, z_2, z_3, z_4) = \int d \h z_1 \la\mO_1 \mO_2 \mO_{\h h_1}(\h z_1) \rangle \langle \t \mO_{\h h_1} (\h  z_1) \mO_3 \mO_4\ra~,
\ee
where $\t\mO_{\h h_1}$ denotes the shadow of $\mO_{\h h_1}$ and has dimension $\tilde{ \h h}_1 = 1- \h h_1$, 
and $\mO_i$ is shorthand for $\mO_{h_i}(z_i)$: an operator of dimension $h_i$. Starting with the $5$-point partial wave, the $n{+}1$-point partial wave in the comb channel is defined  iteratively in terms of the $n$-point partial wave in the comb channel (see Fig.~\ref{nptF}), 
\be
\Psi_{\h h_1, \ldots, \h h_{n-3}, \h h_{n-2}}^{h_1, \ldots, h_{n}, h_{n+1}}(z_1, \ldots, z_{n+1}) = \int d\h z_{n-2}\, \Psi_{\h h_1, \ldots, \h h_{n-3}}^{h_1, \ldots, h_{n-1}, \h h_{n-2}}(z_1, \ldots, z_{n-1}, \h z_{n-2})\, \la \tilde \mO_{\h h_{n-2}}(\h z_{n-2}) \mO_n \mO_{n+1}\ra~.
\ee
The comb channel involves, at each iteration, gluing a three-point function to the end of the partial wave. In some different channel, one would, at some stage, glue the three point function somewhere else. 
The $n{+}1$ point partial wave involves $n{-}2$ integrals of a product of $n{-}1$ conformal $3$-point functions. For instance, applying this equation for $n=4$, and making use of (\ref{4ptCPW}), gives the 5-point partial wave,
\be \label{5ptCPW}
\Psi^{h_1, h_2, h_3, h_4, h_5}_{\h h_1, \h h_2} (z_1, z_2, z_3, z_4, z_5)= \int d \h z_1 d\h z_2\, \la \mO_1 \mO_2 \mO_{\h h_1}(\h z_1)\ra \la \t \mO_{\h h_1}(\h z_1) \mO_3 \mO_{\h h_2}(\h z_2)\ra \la \t \mO_{\h h_2}(\h z_2) \mO_4 \mO_5\ra~.
\ee

The conformal partial waves give a sum of a conformal block and shadow blocks. In particular, the $4$-point partial wave is a sum of two terms: the conformal block for exchange of  an operator of dimension $\h h_1$ and the conformal block for exchange of the shadow operator of dimension $ \tilde{\h h}_1 = 1 - \h h_1$. The $n$-point block is a sum of $2^{n-3}$ terms, accounting for the blocks with exchanged operator dimensions $(\h h_1, \ldots, \h h_{n-3})$ and all possible shadows. 

One can view the insertion of $\int d \h z\, |\mO_{\h h}(\h z)\ra \la \tilde \mO_{\h h}(\h z)|$ as a conformally invariant projector onto the exchange of the operator $\mO_{\h h}$, or its shadow, $\tilde{\mO}_{\h h}$. The shadow formalism \cite{Ferrara1972,Ferrara:1972xe,Ferrara:1972uq} is a useful way of computing conformal blocks \cite{SimmonsDuffin:2012uy}, and one we will exploit.

\subsubsection{Result for $n$-point conformal blocks}
Later in the section, we will compute the $n$-point conformal block in the comb channel, and find it to be, 
\bml 
g_{\h h_1, \ldots, \h h_{n-3}}^{h_1, \ldots, h_n}(\chi_1, \ldots, \chi_{n-3})
=\prod_{i=1}^{n-3} \chi_i^{\h h_i}\\
\, \, \FK{5}{3}{h_1 {+}\h h_1{ -} h_2\kk, \h h_1{ +} \h h_2 {- }h_3\kk, \ldots\kk, \h h_{n-4} {+ }\h h_{n-3} {- }h_{n-2}\kk, \h h_{n-3} {+ }h_n{ -} h_{n-1}}{2 \h h_1\kk, \ldots, 2 \h h_{n-3}}{\chi_1\kk,\!\! \ldots\kk,\! \chi_{n-3}}~, \label{nptg0}
\end{multline} 
where $F_K$ is a multivariable hypergeometric function, which we will refer to as the comb function. The comb function is defined by the sum, 
\bml \label{CombDef}
\FK{5}{3}{a_1\kk, b_1\kk,\!\ldots\kk, b_{k-1}\kk, a_2}{c_1\kk, \!\ldots\kk, c_k}{x_1\kk,\!\!\ldots\kk, \!x_k} \\
= \sum_{n_1, \ldots, n_k=0}^{\infty} \!\!\frac{(a_1)_{n_1} (b_1)_{n_1 + n_2} (b_2) _{n_2+n_3} \cdots (b_{k-1})_{n_{k-1} +n_k} (a_2)_{n_k}}{(c_1)_{n_1}\cdots (c_k)_{n_k}}\frac{x_1^{n_1}}{n_1!}\! \cdots\!\frac{x_k^{n_k}}{n_k!}~,
\end{multline}
where $(a)_n =  \G(a+n)/\G(a)$ is the Pochhammer symbol. 
Some properties of the comb function are derived in Appendix.~\ref{appFK}.~\footnote{The comb function recently appeared in \cite{Belitsky} in the computation of $SL_2$ blocks for null polygon Wilson loops \cite{Alday:2010ku, Sever:2011pc}. There may be a direct mapping between that result and ours, which would be interesting to understand. }

In the special case of two variables, the comb function reduces to the Appell function $F_2$. It will be convenient to define the one variable comb function to be the standard hypergeometric function ${}_2 F_1$. The conformal blocks (\ref{nptg0}) are for the case that all the cross ratios are between zero and one, $0<\chi_i<1$. This occurs if one takes all the positions to be ordered or antiordered. We will assume that they are ordered, $z_1>z_2> \ldots>z_n$. 

A nice property of the conformal blocks in the comb channel is that they have a $\mathcal{Z}_2$ symmetry:  Fig.~\ref{nptF} can be read either from left to right, or from right to left.~\footnote{For the partial waves, one should draw the diagram with arrows on the internal lines. The arrows would all point to the right, with the convention that an arrow leaves an operator and enter the shadow of an operator. The arrows break the symmetry, but only by trivial shadow transform factors that can be accounted for.} They also have a  shift symmetry. Both of these symmetries are reflected in the formula. 

\begin{figure}
\centering
\subfloat[]{
\includegraphics[width=2in]{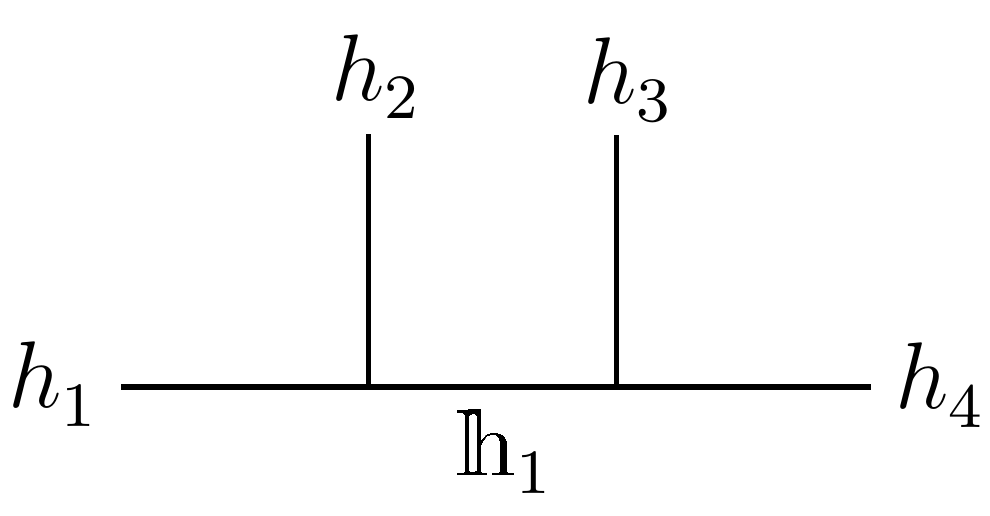}}\, \, \ \ \ \   \ \ \ \ \ \ \ \ \ 
\subfloat[]{
\includegraphics[width=2.6in]{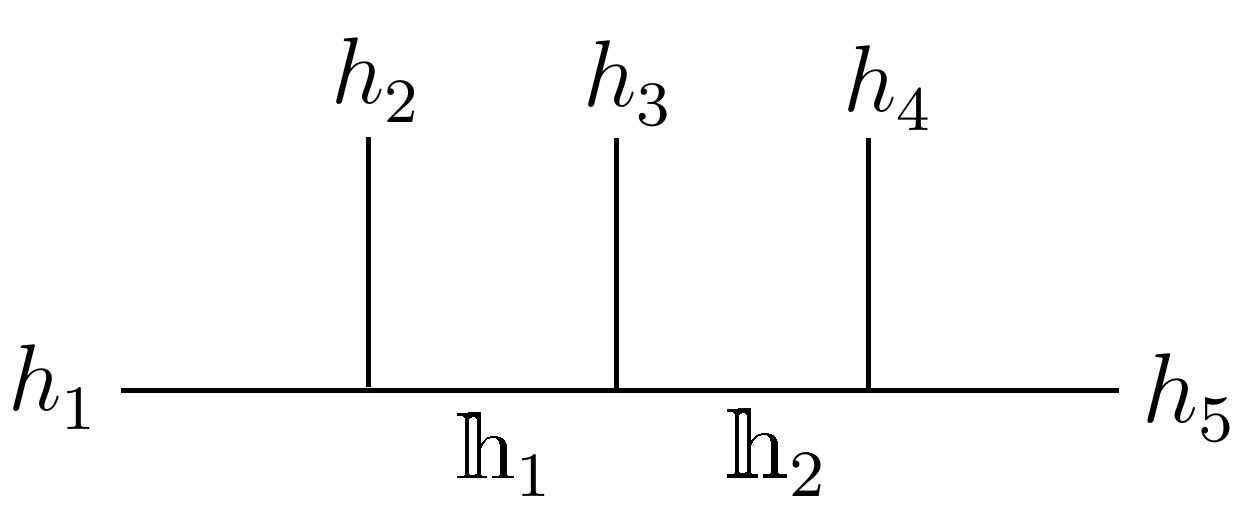}} \ \ \ \ \ 
\subfloat[]{
\includegraphics[width=3in]{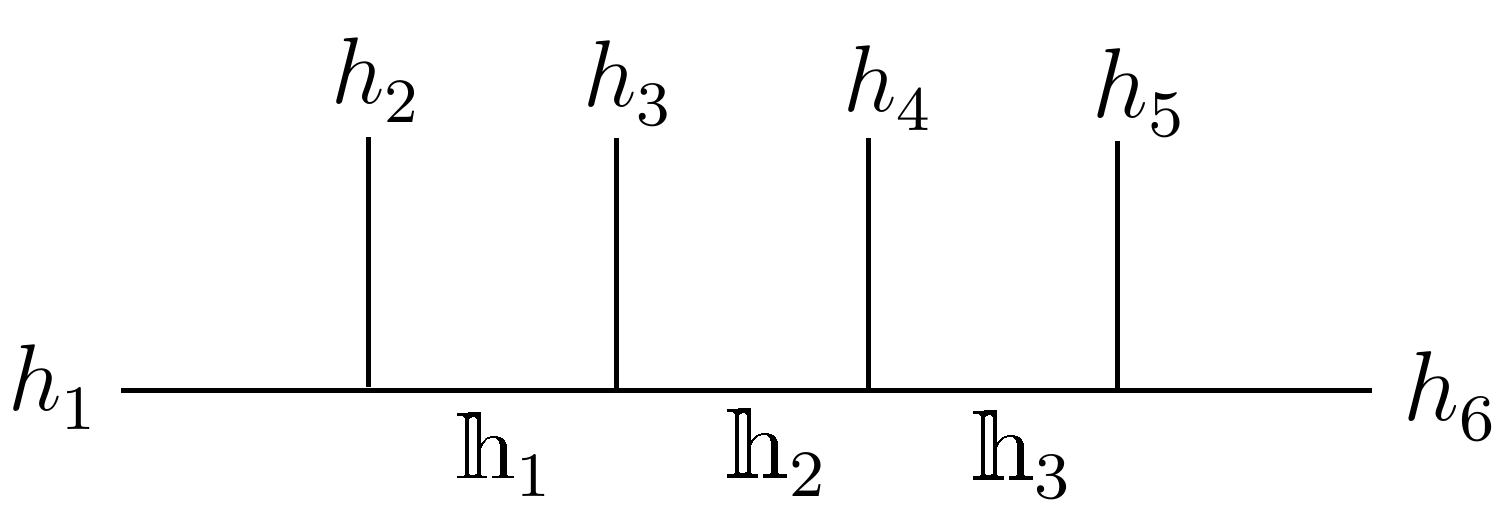}}
\caption{The (a) $4$-point (b) $5$-point (c) $6$-point conformal partial waves.  These are special cases of Fig.~\ref{nptF} for $n=4,5,6$, respectively. } \label{4ptF}
\end{figure}

\subsubsection{Conformal blocks for small $n$}
Let us write out the conformal blocks (\ref{nptg0}) for some small values of $n$. For the $4$-point block, $n=4$, we recover the usual conformal block, see Fig.~\ref{4ptF}(a),~\footnote{Our 4-point block is written in a form that is slight different from  the standard form, due to our choice of leg factor. If one applies the identity for ${}_2 F_1$ in (\ref{2F1Iden}), 
then this gives the standard form of the block.}
\be \label{4ptg}
g^{h_1, h_2, h_3, h_4}_{\h h_1} = \chi_1^{\h h_1}\, \FK{2}{1}{ \h h_1 {+} h_{12}\kk, \h h_1{ -} h_{34}}{2 \h h_1}{\chi_1}~.
\ee
For the $5$-point block, see Fig.~\ref{4ptF}(b), we have, 
\be \label{5ptg}
g^{h_1, h_2, h_3, h_4, h_5}_{\h h_1, \h h_2} = \chi_1^{\h h_1}  \chi_2^{\h h_2}\, \FK{3}{2}{\h h_1 {+} h_{12}\kk, \h h_1 {+} \h h_2 { -} h_3\kk, \h h_2 -{} h_{45}}{2\h h_1\kk, 2 \h h_2}{\chi_1\kk, \chi_2}~,
\ee
and for the $6$-point block, see Fig.~\ref{4ptF}(c), we have, 
\be \label{6ptg}
g^{h_1, h_2, h_3, h_4, h_5, h_6}_{\h h_1, \h h_2, \h h_3} = \chi_1^{\h h_1}  \chi_2^{\h h_2} \chi_3^{\h h_3}\, \FK{4}{3}{\h h_1 {+} h_{12}\kk, \h h_1{+} \h h_2{-}h_3\kk, \h h_2 {+} \h h_3 {-} h_4\kk, \h h_3 {-} h_{56}}{2\h h_1\kk, 2\h h_2\kk, 2 \h h_3}{\chi_1\kk, \chi_2\kk, \chi_3}~,
\ee
where the cross-ratios are (\ref{GDEF}), 
\be
\chi_1 = \frac{z_{12} z_{34}}{z_{13} z_{24}}~, \ \ \ \chi_2 = \frac{z_{23} z_{45}}{z_{24} z_{35}}~, \ \ \ \chi_3 = \frac{ z_{34} z_{56}}{z_{35} z_{46}}~,
\ee
and we have used short-hand $h_{i j} \equiv h_i - h_j$. 

Let us also write out the leg factors (\ref{Leg}) for some small values of $n$. 
For $n=3$ the leg factor is just a conformal three-point function, 
\be
\mL^{h_1, h_2, h_3}=  \(\frac{z_{23}}{z_{12} z_{13}}\)^{h_1}\(\frac{z_{13}}{z_{12} z_{23}}\)^{h_2}\, \(\frac{z_{12}}{z_{13}z_{23}}\)^{h_3} = \frac{1}{z_{12}^{h_1 + h_2 - h_3} z_{13}^{h_1 + h_3 - h_2} z_{23}^{h_2 + h_3 - h_1}}~.
\ee
For $n=4$ we may write the leg factor as,
\be \label{L4}
\mL^{h_1, h_2, h_3, h_4} = \frac{1}{z_{12}^{h_1+h_2} z_{34}^{h_3 + h_4}} \(\frac{z_{23}}{z_{13}}\)^{h_{12}} \(\frac{z_{24}}{z_{23}}\)^{h_{34}}~,
\ee
and for $n=5$ we may write the leg factor as,
\be \label{L5}
\mL^{h_1, h_2, h_3, h_4, h_5} = \frac{1}{z_{12}^{ h_1 +  h_2} z_{34}^{h_3} z_{45}^{h_4 + h_5}} \(\frac{z_{23}}{z_{13}}\)^{h_{12}} \(\frac{z_{24}}{z_{23}}\)^{h_3} \(\frac{z_{35}}{z_{34}}\)^{h_{45}}~.
\ee
From the definition of the leg factors (\ref{Leg}) one can trivially relate the $n{+}1$ point leg factor to the $n$ point leg factor, 
\be \label{Lrel}
\mL^{h_1, \ldots,  h_{n+1}}(z_1, \ldots, z_{n+1}) = \chi_{n-2}^{-h_n} \(\frac{z_{n- 1, n}}{z_{n-1, n+1} z_{n, n+1}}\)^{h_{n+1}}\mL^{h_1, \ldots,  h_{n}}(z_1, \ldots, z_{n})~.
\ee

This completes our summary of the results. 
The rest of the section is devoted to deriving the conformal blocks (\ref{nptg0}). 

\subsection{Four-point block} \label{4ptSec}
We start by recalling the standard four-point conformal block, see Fig.~\ref{4ptF}(a). We will find it by computing the integral defining the conformal partial wave. 
The  definition of the conformal partial wave is, (\ref{4ptCPW}), 
\be  \label{4ptCPW2}
\Psi_{\h h_1}^{h_1, h_2, h_3, h_4}=\int d \h z_1\frac{|z_{12}|^{-h_1 - h_2 + \h h_1} |z_{34}|^{-h_3- h_4 +1-\h h_1}}{|\h z_1 - z_1|^{\h h_1 + h_{12}} | \h z_1 - z_2|^{\h h_1 - h_{12}}|\h z_1 - z_3|^{1- \h h_1 + h_{34}} |\h z_1 - z_4|^{1 - \h h_1 - h_{34}}}~.
\ee
We do a change of variables, $\h z_1 \rightarrow {\h z_1}^{-1} + z_1$, 
\be \label{4ptCPW3}
\Psi_{\h h_1}^{h_1, h_2, h_3, h_4}=\int d \h z_1
\frac{|z_{12}|^{-2 h_2 } 
 |z_{34}|^{-h_3- h_4 +1-\h h_1} |z_{13}|^{-1 + \h h_1 - h_{34}} |z_{14}|^{-1 + \h h_1 + h_{34}}}{|\h z_1 - z_{21}^{-1}|^{\h h_1 - h_{12}}|\h z_1 - z_{31}^{-1}|^{1- \h h_1 +h_{34}}|\h z_1 - z_{41}^{-1}|^{1 - \h h_1 - h_{34}}}~.
\ee
The integral has now simplified: instead of $\h z_1$ colliding with four points, it now only collides with three points. In particular,  after the change of variables we do not have a term in the integrand involving $|\h z_1|$ to some power. This is because the sum of the exponents in the denominator of the integrand in  (\ref{4ptCPW2}) is equal to two.  This is a result of conformal invariance. Namely, for (\ref{4ptCPW2}) to transform correctly under inversion of the points $z_i \rightarrow z_i^{-1}$, it is necessary that the sum of the exponents in the denominator of (\ref{4ptCPW2}) be equal to two (as one can see by inverting the integration point $\h z_1 \rightarrow \h z_1^{-1}$). Performing a further change of variables on (\ref{4ptCPW3}),  $\h z_1 \rightarrow z_{41}^{-1} - \h z_1 (z_{41}^{-1} - z_{21}^{-1})$, gives, 
\be \label{CPW4}
\Psi_{\h h_1}^{h_1, h_2, h_3, h_4}=\frac{1}{|z_{12}|^{h_1 + h_2} |z_{34}|^{h_3+ h_4}}\Big|\frac{z_{24}}{z_{14}}\Big|^{h_{12}}\Big|\frac{z_{14}}{z_{13}}\Big|^{h_{34}}\int d \h z_1 \frac{|\chi_1|^{1-\h h_1}}{|\h z_1|^{1-\h h_1-h_{34}}|1-\h z_1|^{\h h_1 - h_{12}}|\chi_1 - \h z_1 |^{1-\h h_1 + h_{34}}}
\ee
Let us now assume that $0<\chi_1<1$. The integrand is analytic in four separate regions of $\h z_1$, which are: $(-\infty, 0), (0, \chi_1), ( \chi_1,1), (1, \infty)$. We should do the integral in each of the four regions. In the region $0<\h z_1<\chi_1$, the integral is proportional to, 
\be \label{oneReg}
\chi_1^{\h h_1}\, \FK{2}{1}{ \h h_1 {-} h_{12}\kk, \h h_1 {+} h_{34}}{2 \h h_1}{\chi_1}~,
\ee
as one can see from the definition of the hypergeometric function, (\ref{2F1Int}) in Appendix.~\ref{appFK}, if one sends $\h z_1 \rightarrow \chi_1 \h z_1$ (recall that we have extended notation to let $F_K$ in the one variable case denote ${}_2 F_1$). By a hypergeometric identity,
\be \label{2F1Iden}
 \FK{2}{1}{ \h h_1 {-} h_{12}\kk, \h h_1 {+} h_{34}}{2 \h h_1}{\chi_1} = (1-\chi_1)^{h_{12} - h_{34}} \FK{2}{1}{ \h h_1 {+} h_{12}\kk, \h h_1 {-} h_{34}}{2 \h h_1}{\chi_1}~,
 \ee
and so we have obtained the 4-point conformal block, (\ref{4ptg}, \ref{L4}).

In the other three regions, the integral can also be evaluated  by performing  simple variable changes. The result for each of the regions will be a superposition of the conformal block and the shadow block. In fact, this is guaranteed: once we evaluate the integral in the region $0<\h z_1<\chi_1$ and obtain (\ref{oneReg}), we know that the conformal partial wave must be a solution of the second order differential equation obeyed by the hypergeometric function. This has two solutions, corresponding to the conformal block and the shadow block. Thus, the partial wave must be a superposition of these two solutions. We have learned a lesson that will be useful later on: to find the conformal block, it is sufficient to evaluate the integral for the conformal partial wave (\ref{CPW4}) in any of the the four regions.

To go from the definition of  the partial wave (\ref{4ptCPW2}) to the form in (\ref{CPW4}), we did two simple changes of integration variables. Combining them, the transformation is, 
\be
\h z_1 \rightarrow z_1 + \frac{1}{z_{41}^{-1} - \h z_1 (z_{41}^{-1} - z_{21}^{-1})} = \frac{z_1 z_{24} \h z_1 + z_4 z_{12}}{z_{24} \h z_1 + z_{12}}~.
\ee
This transformation sends the four points $\h z_1 = (z_4, z_3, z_2, z_1) \leftarrow \h z_1 = (0, \chi_1, 1, \pm \infty)$.  
Of course, the reason that this change of variables simplified the integral is $SL_2$ symmetry. We could have equivalently taken the original integral (\ref{4ptCPW2}) and set $z_4 = 0, z_3 = \chi_1, z_2=1, z_1 = \infty$. 

To summarize: one method for computing the conformal block is to start with the definition of the partial wave (\ref{4ptCPW}) and to evaluate the integral. One can simplify the problem by restricting the integration range to lie in any of the four regions in between the singularities where  points collide:  $\h z_1 \subset (z_4, z_3)$ or  $\h z_1 \subset (z_3, z_2)$ or  $\h z_1 \subset  (z_2, z_1)$ or  $\h z_1 \subset (- \infty, z_4) \cup (z_1, \infty) $, where we have assumed $z_1> z_2>z_3>z_4$. The integral from any one of these four regions gives some combination of the  block and the shadow block. Distinguishing between the two is trivial, and so this is enough to find the block.~\footnote{ If one is interested in the conformal partial wave, then this can easily be established from knowledge of the conformal block. The partial wave is a sum of a block and a shadow block, with some coefficients which can be established by evaluating the integral in the OPE limit, see e.g. \cite{GR4, Karateev:2018oml}. The same applies to higher point partial waves as well.}

\subsection{Five-point block} \label{5ptSec}

\subsubsection*{Direct evaluation}
We start with the definition of the 5-point conformal partial wave, (\ref{5ptCPW}),
and change integration variables, $\h z_1 \rightarrow z_3 -{ \h z_1}^{-1}$ and $\h z_2 \rightarrow z_3 -{ \h z_2}^{-1}$ to obtain, 
\bea
\Psi^{h_1, h_2, h_3, h_4, h_5}_{\h h_1, \h h_2} &=& \frac{|z_{12}|^{\h h_1 - h_1 - h_2} |z_{45}|^{1- \h h_2 - h_4 - h_5}}{|z_{13}|^{\h h_1 + h_{12}} |z_{23}|^{\h h_1 - h_{12}} |z_{43}|^{1- \h h_2 + h_{45}}|z_{53}|^{1- \h h_2 - h_{45}}} \\ \nonumber
&& \int d\h z_1 d\h z_2\frac{\, |\h z_1 - \h z_2|^{\h h_1 - \h h_2 +h_3 - 1}\, }{|\h z_1 + z_{13}^{-1}|^{\h h_1 + h_{12}}|\h z_1 + z_{23}^{-1}|^{\h h_1 - h_{12}} |\h z_2 + z_{43}^{-1}|^{1-\h h_2 + h_{45}} |\h z_2+ z_{53}^{-1}|^{1- \h h_2 -h_{45}}}
\eea
Performing a further change of variables, $\h z_{1} \rightarrow \h z_1 (z_{23}^{-1} - z_{13}^{-1}) - z_{23}^{-1}$ and $\h z_2 \rightarrow \h z_2 ( z_{43}^{-1} - z_{53}^{-1}) - z_{43}^{-1}$, we get \cite{GR4},
\be
\Psi^{h_1, h_2, h_3, h_4, h_5}_{\h h_1, \h h_2} = \mL^{h_1,  h_2, h_3, h_4, h_5}\,\int d \h z_1 d \h z_2 \frac{ |\chi_1|^{1-\h h_1} |\chi_2|^{\h h_2} \, \, |1 - \h z_1 \chi_1 - \h z_2 \chi_2|^{\h h_1 - \h h_2 + h_3 -1}}{|\h z_1|^{\h h_1 - h_{12}} |1 - \h z_1|^{\h h_1 + h_{12}} |\h z_2|^{1- \h h_2 + h_{45}}|1- \h z_2|^{1-\h h_2 - h_{45}}}~.
\ee
The portion of the integral from the region of integration $0< \h z_1<1$ and $0 < \h z_2<1$ is an Appell function (\ref{AppellInt}), 
\be
\chi_1^{1-\h h_1} \chi_2^{\h h_2}\, \FK{3}{2}{1 {-} \h h_1{ +} h_{12}\kk,1{-}\h h_1 {+ }\h h_2{-}h_3\kk,  \h h_2 {-} h_{45}}{2 {-} 2\h h_1\kk, \h h_2}{\chi_1\kk, \chi_2}~.
\ee
Taking the shadow with respect to $\h h_1$, by sending $\h h_1 \rightarrow 1- \h h_1$, we obtain the 5-point conformal block (\ref{5ptg}).

The derivation we just gave of the $5$-point conformal block relied on starting with the definition of the conformal partial wave, and  finding an appropriate change of integration variables that gives an integrand depending only on the conformal cross-ratios $\chi_1, \chi_2$. Obtaining an integrand that depends only on the cross-ratios is not difficult: one can, without loss of generality, take the definition of the conformal partial wave (\ref{5ptCPW}) and write it as the leg factor $\mL^{h_1, \ldots, h_5}$ times some function of the two cross-ratios, and then make some choice for the external points, such as $z_1 = \infty$, $z_2 = 1$, $z_5=0$, and express the remaining two points $z_3, z_4$ appearing in the partial wave in terms of $\chi_1, \chi_2$. What is slightly more challenging, if one does not make the best choice, is recognizing the resulting integral as some special function. In the case of the $5$-point block, the special function one is looking for is a two variable function, the simplest ones are Appell functions, and there are only four kinds of Appell functions. As a result, one has a fairly good sense of what the answer might be.

 For higher point blocks, it will be harder to find  a good change of variables that turns the integral into something we can recognize. In part, this is because there are a large number of multivariable hypergeometric functions, when the number of variables is greater than two. The key step in our derivation of  the $n$-point block  will be to make use of the result for the $n{-}1$ point block, as input in computing the $n$-point block. 
 
 In order to illustrate the method, we will now redo the computation of the $5$-point block, in a way that makes use of the $4$-point block. 
 
 \subsubsection*{From 4-point to 5-point blocks}
 We start with the definition of the 5-point conformal partial wave in terms of the 4-point conformal partial wave, 
\be \label{5ptCPW0}
\Psi_{\h h_1, \h h_2}^{h_1, h_2, h_3, h_4, h_5} (z_1,z_2, z_3, z_4, z_5) =\int d \h z_2\, \Psi_{\h h_1}^{h_1, h_2, h_3, \h h_2}(z_1, z_2, z_3, \h z_2)\la \tilde \mO_{\h h_2}(\h z_2) \mO_4 \mO_5\ra~.
\ee
We will assume that the external positions are time ordered, $z_1> z_2>z_3>z_4>z_5$. The integral, as stated, requires us to integrate over all $- \infty<\h z_2<\infty$. However, as we discussed previously, for finding the conformal block, it is enough to only look at the integral in one of the regions where it is analytic. We may therefore consider just the portion of the integral coming from $z_3> \h z_2> z_4$. We may also replace the 4-point partial wave by the 4-point conformal block, since we are not interested in the shadow block. Thus,~\footnote{Our usage of $A \supset B$  denotes $A$ containing $B$, up to some prefactor. We know that the term on the right in   (\ref{5ptCPW2})  is a linear combination of the conformal block and its shadows. These all appear in the conformal partial wave, but not with the coefficients produced from the term on the right.}
\be \label{5ptCPW2}
\Psi_{\h h_1, \h h_2}^{h_1, h_2, h_3, h_4, h_5} (z_1,z_2, z_3, z_4, z_5) \supset \int_{z_4}^{z_3} d \h z_2\, G_{\h h_1}^{h_1, h_2, h_3, \h h_2}(z_1, z_2, z_3, \h z_2)\la \tilde \mO_{\h h_2}(\h z_2) \mO_4 \mO_5\ra~.
\ee
We write the 4-point conformal block appearing above in terms of the leg factor and the bare conformal block, 
\be
G_{\h h_1}^{h_1, h_2, h_3, \h h_2}(z_1, z_2, z_3, \h z_2) = \mL^{h_1, h_2, h_3, \h h_2}(z_1, z_2, z_3, \h z_2)\, g_{\h h_1}^{h_1, h_2, h_3, \h h_2}(\rho)~, \ \ \ \ \rho = \frac{z_{12} (z_3 - \h z_2)}{z_{13} (z_2 - \h z_2)}~,
\ee
where the leg factor is (\ref{Lrel}), 
\be
 \mL^{h_1, h_2, h_3, \h h_2}(z_1, z_2, z_3, \h z_2) =  \mL^{h_1, h_2, h_3}(z_1, z_2, z_3)\, \rho^{-h_3} \(\frac{z_{23}}{(z_2 - \h z_2) (z_3- \h z_2)}\)^{\h h_2}~,
 \ee
 while the bare conformal block is (\ref{4ptg}), 
 \be
  g_{\h h_1}^{h_1, h_2, h_3, \h h_2}(\rho) = \sum_{m=0}^{\infty} \frac{(\h h_1 + h_{12})_m (\h h_1+ \h h_2 - h_3)_m}{(2\h h_1)_m}\frac{\rho^{m+\h h_1}}{m!} ~.
 \ee
 The restriction on the integration range we made in (\ref{5ptCPW2}), that $\h z_2< z_3$, implies $0<\rho<1$, and allows us to use the 4-point conformal block in the form above.  Inserting these terms gives, 
 \bml 
\Psi_{\h h_1, \h h_2}^{h_1, h_2, h_3, h_4, h_5} (z_1,z_2, z_3, z_4, z_5) \supset \mL^{h_1, h_2, h_3}(z_1, z_2, z_3)\sum_{m=0}^{\infty} \frac{(\h h_1 + h_{12})_m (\h h_1+ \h h_2 - h_3)_m}{(2\h h_1)_m m!}  \(\frac{z_{12}}{z_{13}}\)^{m+\h h_1 - h_3}\\ z_{23}^{m+\h h_1 +h_3}
 \int_{z_4}^{z_3} d \h z_2\, \la \mO_{\h h_1 + m}(z_2) \mO_3 \mO_{\h h_2}(\h z_2)\ra\la \tilde \mO_{\h h_2}(\h z_2) \mO_4 \mO_5\ra~,
\end{multline}
where the first three-point function in the integral emerged from the powers of $\h z_2 - z_3$ and $\h z_2 - z_2$ that occurred. We recognize the integral here as the same one defining the conformal partial wave, $\Psi^{\h h_1 + m, h_3, h_4, h_5}_{\h h_2}(z_2, z_3, z_4, z_5)$. As discussed in Sec.~\ref{4ptSec}, restricting the integration range to $z_4 < \h z_2< z_3$ will still give rise to a linear combination of the  block and its shadow. We may therefore replace the integral above with the conformal block $G^{\h h_1 + m, h_3, h_4, h_5}_{\h h_2}(z_2, z_3, z_4, z_5)$, which we write in terms of the leg factor and the bare block, to get
 \bml 
\Psi_{\h h_1, \h h_2}^{h_1, h_2, h_3, h_4, h_5} \supset \mL^{h_1, h_2, h_3}(z_1, z_2, z_3)\sum_{m=0}^{\infty} \frac{(\h h_1 + h_{12})_m (\h h_1+ \h h_2 - h_3)_m}{(2\h h_1)_m m!}  \(\frac{z_{12}}{z_{13}}\)^{m+\h h_1 - h_3} z_{23}^{m+\h h_1 +h_3}\\
\mL^{\h h_1 + m, h_3, h_4, h_5}(z_2, z_3, z_4, z_5)\, g_{\h h_2}^{\h h_1+m, h_3, h_4, h_5}(\chi_2)~.
\end{multline}
Writing out the leg factors, we reorganize this as, 
 \be \nonumber
\Psi_{\h h_1, \h h_2}^{h_1, h_2, h_3, h_4, h_5} \supset \mL^{h_1, h_2, h_3, h_4, h_5}(z_1, z_2, z_3, z_4, z_5)\sum_{m=0}^{\infty} \frac{(\h h_1 + h_{12})_m (\h h_1+ \h h_2 - h_3)_m}{(2\h h_1)_m\, m!} \chi_1^{m+\h h_1}
\, g_{\h h_2}^{\h h_1+m, h_3, h_4, h_5}(\chi_2)~.
\ee
We identify the sum as the bare 5-point conformal block. Inserting the 4-point conformal block (\ref{4ptg}), written as a sum, we thus have, 
\be 
g_{\h h_1, \h h_2}^{h_1, h_2, h_3, h_4, h_5} =\chi_1^{\h h_1} \chi_2^{\h h_2} \sum_{m, n=0}^{\infty} \frac{(\h h_1 + h_{12})_m (\h h_1+ \h h_2 - h_3)_m}{(2\h h_1)_m } \frac{(\h h_2 + \h h_1+m -h_3)_n(\h h_2 - h_{45})_n}{(2\h h_2)_n}\frac{\chi_1^{m}}{m!}\frac{\chi_2^n}{n!}~.
\ee
Noting that $(\h h_1 + \h h_2+m -h_3)_n = (\h h_1+ \h h_2 - h_3)_{n+m}/(\h h_1 + \h h_2 - h_3)_m$, we see that this sum is just the Appell function $F_2$, and we thus get the 5-point conformal block stated earlier (\ref{5ptg}).

\subsection{$n$-point block} \label{nptSec}

To obtain the $n{+}1$ point block from the $n$ point block we use a similar procedure as the one used in the previous section for getting the 5-point block from the 4-point block. Namely, we start with the definition of the $n{+}1$ point conformal partial wave as an $n$-point partial wave glued to a 3-point function. We use this expression, with the range of integration restricted, and insert the $n$-point block, rewritten as a sum involving an $n{-}1$ point block (so as to strip off the last cross-ratio), and recognize the integral to be the one that gives rise to a $4$-point conformal block (involving the last $4$ points). The sum then defines the $n{+}1$ point block. After doing this for the $6$-point and $7$-point blocks, the pattern becomes clear, allowing us to guess  the $n$ point block,  (\ref{nptg0}). We will show by induction that this guess is correct.

We start with the definition of the conformal partial wave in the comb channel,
\be
\Psi_{\h h_1, \ldots, \h h_{n-3}, \h h_{n-2}}^{h_1, \ldots, h_{n}, h_{n+1}}(z_1,\! \ldots\!, z_{n+1}) = \int d\h z_{n-2}\, \Psi_{\h h_1, \ldots, \h h_{n-3}}^{h_1, \ldots, h_{n-1}, \h h_{n-2}}(z_1,\! \ldots\!, z_{n-1}, \h z_{n-2})\, \la \tilde \mO_{\h h_{n-2}}(\h z_{n-2}) \mO_n \mO_{n+1}\ra~.
\ee
As we did in the derivation of the 5-point block, we restrict the range of integration on the right-hand side, and in addition replace the partial wave with the block, 
\be
\Psi_{\h h_1, \ldots, \h h_{n-3}, \h h_{n-2}}^{h_1, \ldots, h_{n}, h_{n+1}}(z_1,\! \ldots\!, z_{n+1}) \supset \int_{z_{n}}^{z_{n-1}} d\h z_{n-2}\, G_{\h h_1, \ldots, \h h_{n-3}}^{h_1, \ldots, h_{n-1}, \h h_{n-2}}(z_1, \!\ldots\!, z_{n-1}, \h z_{n-2})\, \la \tilde \mO_{\h h_{n-2}}(\h z_{n-2}) \mO_n \mO_{n+1}\ra
\ee
We insert the conformal block appearing on the right, written as a leg factor times the bare block,
\be \nonumber
G_{\h h_1, \ldots, \h h_{n-3}}^{h_1, \ldots, h_{n-1}, \h h_{n-2}}(z_1,\! \ldots\!, z_{n-1}, \h z_{n-2}) =\mL^{h_1, \ldots, h_{n-1}, \h h_{n-2}}(z_1,\! \ldots\!, z_{n-1}, \h z_{n-2})\, g_{\h h_1, \ldots, \h h_{n-3}}^{h_1, \ldots, h_{n-1}, \h h_{n-2}}(\chi_1, \! \ldots\!, \chi_{n-4}, \rho)~,
\ee
where the cross-ratio between the last 4 points in the block is,
\be
\rho = \frac{z_{n-3, n-2} (z_{n-1} - \h z_{n-2})}{z_{n-3, n-1} (z_{n-2} - \h z_{n-2})}~,
\ee
and where the leg factor is, upon using the relation between leg factors (\ref{Lrel}), 
\be \nonumber
\mL^{h_1, \ldots, h_{n-1}, \h h_{n-2}}(z_1,\! \ldots\!, z_{n-1}, \h z_{n-2}) = \mL^{h_1, \!\ldots\!, h_{n-1}}(z_1,\! \ldots\!, z_{n-1}) \(\frac{z_{n-2, n-1}}{(z_{n-2} - \h z_{n-2})(z_{n-1} - \h z_{n-2})}\)^{\h h_{n-2}} \!\!\rho^{- h_{n-1}}~,
\ee
and where the block is written as, using the relation between bare conformal blocks (\ref{grel}),
\bml  \nn
g_{\h h_1, \ldots, \h h_{n-3}}^{h_1, \ldots, h_{n-1}, \h h_{n-2}}(\chi_1, \ldots, \chi_{n-4}, \rho) \\
= \rho^{\h h_{n-3}}
\sum_{m=0}^{\infty} \frac{(\h h_{n-4} {+} \h h_{n-3} {-} h_{n-2})_m (\h h_{n-3}{+} \h h_{n-2} { -} h_{n-1})_m}{(2 \h h_{n-3})_m}
\, \frac{\rho^{m}}{m!}\, g_{\h h_1, \ldots, \h h_{n-4}}^{h_1, \ldots, h_{n-2}, \h h_{n-3}+m}(\chi_1, \ldots, \chi_{n-4})~.
\end{multline}
Thus we have,
\bml
\Psi_{\h h_1, \ldots, \h h_{n-3}, \h h_{n-2}}^{h_1, \ldots, h_{n}, h_{n+1}}(z_1, \ldots, z_{n+1}) \supset  \mL^{h_1, \ldots, h_{n-1}}(z_1, \ldots, z_{n-1}) \sum_{m=0}^{\infty} \frac{(\h h_{n-4} {+} \h h_{n-3} {-} h_{n-2})_m (\h h_{n-3} {+} \h h_{n-2} {-} h_{n-1})_m}{(2 \h h_{n-3})_m\, m!}\\ g_{\h h_1, \ldots, \h h_{n-4}}^{h_1, \ldots, h_{n-2}, \h h_{n-3}+m}(\chi_1, \ldots, \chi_{n-4})\, z_{n-2, n-1}^{h_{n-1} + \h h_{n-3} + m}\, \(\frac{z_{n-3, n-2}}{ z_{n-3, n-1}}\)^{- h_{n-1}+ \h h_{n-3} + m}\\
\int_{z_n}^{z_{n-1}} d\h z_{n-2}\la \mO_{\h h_{n-3}+m}(z_{n-2}) \mO_{n-1} \mO_{\h h_{n-2}}(\h z_{n-2})\ra \la  \tilde \mO_{\h h_{n-2}}(\h z_{n-2}) \mO_n \mO_{n+1}\ra
\end{multline}
We recognize  the integral on the right side is the same one that appears in the definition of the 4-point partial wave, $\Psi_{\h h_{n-2}}^{\h h_{n-3} +m, h_{n-1}, h_n, h_{n+1}}(z_{n-2}, z_{n-1}, z_n, z_{n+1})$, and so we replace this integral by the conformal block $G_{\h h_{n-2}}^{\h h_{n-3} +m, h_{n-1}, h_n, h_{n+1}}(z_{n-2}, z_{n-1}, z_n, z_{n+1})$, which we write as a leg factor times the bare block,
\bml
\!\!\!\Psi_{\h h_1, \ldots, \h h_{n-3}, \h h_{n-2}}^{h_1, \ldots, h_{n}, h_{n+1}}(z_1,\! \ldots\!, z_{n+1}) \supset  \mL^{h_1, \ldots, h_{n-1}}(z_1,\! \ldots\!, z_{n-1}) \sum_{m=0}^{\infty} \frac{(\h h_{n-4} {+} \h h_{n-3} {-} h_{n-2})_m (\h h_{n-3} {+} \h h_{n-2} {-} h_{n-1})_m}{(2 \h h_{n-3})_m\, m!}\\  
z_{n-2, n-1}^{h_{n-1} + \h h_{n-3} + m}\, \(\frac{z_{n-3, n-2}}{ z_{n-3, n-1}}\)^{- h_{n-1}+ \h h_{n-3} + m} \mL^{\h h_{n-3} +m, h_{n-1}, h_n, h_{n+1}}(z_{n-2}, z_{n-1}, z_n, z_{n+1})\\
g_{\h h_1, \ldots, \h h_{n-4}}^{h_1, \ldots, h_{n-2}, \h h_{n-3}+m}(\chi_1, \ldots, \chi_{n-4})\, g_{\h h_{n-2}}^{\h h_{n-3} +m, h_{n-1}, h_n, h_{n+1}}(\chi_{n-2})~.
\end{multline}
We identify the right hand side with the $n{+}1$ point block. Simplifying gives, 
\bml
\!\!\!\!G_{\h h_1, \ldots, \h h_{n-3}, \h h_{n-2}}^{h_1, \ldots, h_{n}, h_{n+1}}(z_1,\! \ldots\!, z_{n+1}) =\mL^{h_1, \ldots, h_{n+1}}(z_1,\! \ldots\!, z_{n+1})\sum_{m=0}^{\infty} \frac{(\h h_{n-4} {+} \h h_{n-3} {-} h_{n-2})_m (\h h_{n-3} {+} \h h_{n-2} {-} h_{n-1})_m}{(2 \h h_{n-3})_m\, m!}\\ \chi_{n-3}^{m+ \h h_{n-3}}  g_{\h h_1, \ldots, \h h_{n-4}}^{h_1, \ldots, h_{n-2}, \h h_{n-3}+m}(\chi_1, \ldots, \chi_{n-4})\, g_{\h h_{n-2}}^{\h h_{n-3} +m, h_{n-1}, h_n, h_{n+1}}(\chi_{n-2})
\end{multline}
Using an identity of the comb function, see (\ref{grel2}) in Appendix.~\ref{appFK}, we recognize the right side is indeed the $n+1$ point block, (\ref{nptg0}).

\subsection{Analysis} \label{Analysis}

\subsubsection{OPE limit}
In this section we check that the conformal blocks behave correctly in the OPE limit, to leading order. 
Consider taking a four-point function $
\la \mO_1 \mO_2 \mO_3 \mO_4\ra$,
and performing the OPE on the last two operator, 
\be
\mO_3 \mO_4 = \sum_{\h h_1} z_{34}^{\h h_1- h_3 - h_4} \mO_{\h h_1}(z_3)+ \ldots~,
\ee
where the dots are the descendants of $\mO_{\h h_1}$. This gives for the four-point function, 
\be \label{4ptOPE}
\la \mO_1 \mO_2 \mO_3 \mO_4\ra = \sum_{\h h_1} z_{34}^{\h h_1- h_3 - h_4} \la \mO_1 \mO_2  \mO_{\h h_1}(z_3)\ra + \ldots~, \ \ \ \ \ \ z_4\rightarrow z_3~.
\ee
Let us check that our conformal blocks behave correctly in the OPE limit. For the 4-point block we take $z_4\rightarrow z_3$, as above. The conformal cross ratio $\chi_1$ goes to zero in this limit, and the behavior of the bare block and the leg factor in this limit is, 
\be
g_{\h h_1}^{h_1, \ldots h_4} \rightarrow \chi_1^{\h h_1}\rightarrow \(\frac{z_{12} z_{34}}{z_{13}z_{23}}\)^{\h h_1}~, \ \ \ \ \mL^{h_1, \ldots, h_4} \rightarrow \frac{1}{z_{12}^{h_1+h_2} z_{34}^{h_3 + h_4}} \(\frac{z_{23}}{z_{13}}\)^{h_{12}}~,\ \ \ \ z_4\rightarrow z_3~,
\ee
where in the denominator of $\chi_1$  in $g_{\h h_1}^{h_1, \ldots h_4}$ we replaced $z_{24}$ with $z_{23}$. Combining the bare block and the leg factor, we have that the block behaves as, 
\be
G^{h_1, \ldots, h_4}_{\h h_1} \rightarrow  z_{34}^{\h h_1 - h_3 - h_4}  \la \mO_1 \mO_2 \mO_{\h h_1}(z_3)\ra~, \ \ \ \ \ z_4 \rightarrow z_3~,
\ee
which is the correct behavior (\ref{4ptOPE}). 

Now, consider performing the OPE on the last two operators in a $n$-point function, 
\be \label{nptOPE}
\la \mO_1 \cdots \mO_{n-2} \mO_{n-1} \mO_n\ra = \sum_{\h h_{n-3}} z_{n-1, n}^{\h h_{n-3} - h_{n-1} - h_n} \la \mO_1 \cdots \mO_{n-2} \mO_{\h h_{n-3}}(z_{n-1})\ra + \ldots~, \ \ \ \ \ \ z_n \rightarrow z_{n-1}~.
\ee
Let us check that our $n$-point conformal block has this behavior in the $z_n \rightarrow z_{n-1}$ limit. The cross ratio $\chi_{n-3} \rightarrow 0$, while all other cross ratios remain finite. Thus, from (\ref{nptg0}), we see that the bare $n$ point block reduces to a bare $n{-}1$ point block, 
\be
g_{\h h_1, \ldots, \h h_{n-3}}^{h_1, \ldots, h_n}(\chi_1, \ldots, \chi_{n-3}) \rightarrow \chi_{n-3}^{\h h_{n-3}}\, g_{\h h_1, \ldots, \h h_{n-4}}^{h_1, \ldots, h_{n-2}, \h h_{n-3}}(\chi_1, \ldots, \chi_{n-4})~, \ \ \ \ z_n \rightarrow z_{n-1}~.
\ee
From the definition of the leg factors (\ref{Leg}), we have the relation, 
\bml \nonumber
\mL^{h_1, \ldots, h_n}(z_1, \!\ldots\!, z_n) \\
= \mL^{h_1, \ldots, h_{n-2}, \h h_{n-3}}(z_1, \!\ldots\!, z_{n-1}) \(\frac{z_{n-2, n-1}}{z_{n-2, n} z_{n-1, n}}\)^{h_n} \(\frac{z_{n-2, n}}{z_{n-2, n-1} z_{n-1, n}}\)^{h_{n-1}} \(\frac{z_{n-3, n-2}}{z_{n-3, n-1} z_{n-2,n-1}}\)^{- \h h_{n-3}}~.
\end{multline}
Using this, and taking the limit $z_n \rightarrow z_{n-1}$, we get, 
\be
\mL^{h_1, \ldots, h_n}(z_1, \ldots, z_n)\chi_{n-3}^{\h h_{n-3}}\rightarrow z_{n-1, n}^{\h h_{n-3} - h_n-h_{n-1}}\, \mL^{h_1, \ldots, h_{n-2}, \h h_{n-3}}(z_1, \!\ldots\!, z_{n-1})~,\ \ \ \ \ z_n \rightarrow z_{n-1}~, 
\ee
and consequently the behavior of the block in the OPE limit is, 
\be
G_{\h h_1, \ldots, \h h_{n-3}}^{h_1, \ldots, h_n}(z_1, \ldots, z_n) \rightarrow  z_{n-1, n}^{\h h_{n-3} - h_n-h_{n-1}} \, G_{\h h_1, \ldots, \h h_{n-4}}^{h_1, \ldots, h_{n-2}, \h h_{n-3}}(z_1, \ldots, z_{n-1})~, \ \ \ \ \ z_n \rightarrow z_{n-1}~, 
\ee
as expected from (\ref{nptOPE}).

\subsubsection{Casimir equations} \label{CasimirSec}
Here we check that the conformal blocks behave correctly as eigenfunctions of the Casimir of the conformal group. In particular, in one dimension the conformal group is $SL_2(R)$, with generators $l_a$ and Casimir $l^2$, 
\be \nonumber
l_{-1} = \partial_z~, \ \ \  l_0 = z \partial_z + \h h~, \ \ \ \ l_1 = z^2 \partial_z + 2\h h\, z~, \ \ \ \ \ l^2= l_0^2-\frac{1}{2} l_{-1} l_1 - \frac{1}{2} l_1 l_{-1} = \h h(\h h-1)~.
\ee
\be
\[l_{-1}, l_0\] = l_{-1}~, \ \ \ \ \[l_{-1}, l_1\] = 2 l_0~, \ \ \ \ \[l_0, l_1\] = l_1~.
\ee
Let us denote the $SL_2(R)$ generator acting on point $z_i$ by $l^{(i)}_a$. We define the Casimir acting on two points, $z_i$ and $z_j$, to be, 
\be
C(i, j) = (l^{(i)} + l^{(j)})^2\equiv (l_{0}^{(i)}+l_{0}^{(j)})^2 -\frac{1}{2} (l_{-1}^{(i)}+l_{-1}^{(j)})(l_{1}^{(i)}+l_{1}^{(j)})-\frac{1}{2} (l_{1}^{(i)}+l_{1}^{(j)})(l_{-1}^{(i)}+l_{-1}^{(j)})~.
\ee
We  make an analogous definition for more points: for instance, $C(i, j, k) = (l^{(i)} + l^{(j)}+l^{(k)})^2$.

A defining property of the 4-point conformal block is that it is an eigenfunction of the Casimir acting on two points \cite{Dolan:2003hv}, 
\be  \label{4ptCasimir}
\[ C(1, 2) - \h h_1 (\h h_1 - 1)\] G_{\h h_1}^{h_1,\ldots, h_4}(z_1,\! \ldots\!, z_4)  = 0~.
\ee
For the 5-point block, there are two equations \cite{Alkalaev:2015fbw}, 
\be
\[C(1,2) - \h h_1 (\h h_1 - 1)\]G_{\h h_1, \h h_2}^{h_1,\ldots, h_5}(z_1,\! \ldots \!,z_5)  = 0~, \ \ \ \[C(4,5) - \h h_2 (\h h_2 - 1)\]G_{\h h_1, \h h_2}^{h_1,\ldots, h_5}(z_1,\! \ldots \!,z_5)  = 0~.
\ee
For the 6-point block, there are three equations: the two above for the 5-point block, and in addition, 
\be \label{252}
\[C(1, 2,3) - \h h_2 (\h h_2 -1)\]G_{\h h_1, \h h_2, \h h_3}^{h_1,\ldots, h_6}(z_1,\!\ldots\!, z_6)  = 0~.
\ee
In any of these equations, it is equivalent to act with the Casimir on the complimentary set of points. For instance, in the 4-point equation we could act with $C(3,4)$ instead of $C(1,2)$, or in the 5-point equation we could act with $C(1,2,3)$ instead of $C(4,5)$. 
For the $n$-point block, the full set of $n-3$ equations are, 
\be
\[C(1,\ldots, k) - \h h_{k-1}(\h h_{k-1} -1)\] G_{\h h_1, \ldots, \h h_{n-3}}^{h_1, \ldots, h_n}(z_1, \ldots, z_n) = 0~, \ \ \ 2\leq k\leq n{-}2~.
\ee

We may insert into the these equations the decomposition of the conformal block into the leg factor and the bare block (\ref{GDEF}), to then obtain sets of differential equations in terms of the cross-ratios. This is simple to do for low values of $n$, and to then recognize the equations,  and to see that the solutions are what we wrote down before for the conformal blocks. 

This is a good approach for checking that one has the correct conformal blocks, but is less good for actually finding the blocks, for $n\geq 5$, due to the large number of choices one needs to make: the choice of the leg factors, as well as the choice of the cross-ratios. Any (correct) choice will give correct equations, but they may not be in a form in which one can recognize the solution.
In Appendix~\ref{CombDiff} we derive differential equations for the comb function, which are what these Casimir equations become.~\footnote{We checked that the Casimir equations for $n$-point blocks are the same as the differential equations for the comb function, for $n$ up to $6$. It is straightforward to check for larger $n$, but it is not obvious how to make  the match manifest.}

\section{$d$ Dimensions} \label{dd}

In this section we study conformal blocks in general dimension $d$. This is more involved than in one or two dimensions, and we restrict to external operators that are scalars and exchanged operators that are scalars. In Sec.~\ref{4ptd} we review the computation of the $4$-point blocks. In Sec.~\ref{5ptd} we compute the $5$-point blocks.

\subsection{Four-point block, scalar exchange} \label{4ptd}

The four-point conformal partial wave is, 
\be
\Psi^{\Delta_i}_{\Delta}(x_i) = \int d^d x_0\, \la \mO_1 \mO_2 \mO(x_0)\ra \la \tilde \mO(x_0) \mO_3 \mO_4\ra~,
\ee
where $\Delta_i$ are the dimensions of the external operators,  and $\Delta$ is the dimension of the exchanged operator $\mO$. We take all operators to be scalars.~\footnote{It is common to denote the partial wave by $\Psi^{\Delta_i}_{\Delta,J}$ with $J$ denoting the spin of the exchanged operator. We have not written our partial wave in this notation, as $\Psi^{\Delta_i}_{\Delta,0}$, because in the case of the 5-point blocks, the second subscript will denote the dimension of the other exchanged operator. } Here $\tilde \mO$ refers to the shadow of $\mO$, which has dimension $\tilde \D = d- \Delta$. 

Writing out the three-point functions, the integral we must evaluate is, 
\be  \label{4pw}
\Psi^{\Delta_i}_{\Delta}(x_i)=  \int d^d x_0 \frac{X_{12}^{\frac{\Delta - \Delta_1 - \Delta_2}{2}} X_{34}^{\frac{\t \Delta - \Delta_3 - \Delta_4}{2}}}{X_{10}^{\frac{\Delta + \Delta_{12}}{2}} X_{20}^{\frac{\Delta - \Delta_{12}}{2}} X_{30}^{\frac{\t \Delta + \Delta_{34}}{2}} X_{40}^{\frac{\t \Delta - \Delta_{34}}{2}}}~, \  \ \ \ \ \ \  X_{i j} \equiv (x_i- x_j)^2~. 
\ee
This is a 4-point integral, of the form studied in  Appendix.~\ref{MBapp}. Applying (\ref{I4app}) the result is a sum of the conformal block and the shadow block ,
\be
\Psi^{\Delta_i}_{\Delta}(x_i) = K_{\tilde \D}^{\D_3, \D_4}G_{\Delta}^{\Delta_i}(x_i)+ K_{\D}^{\D_1, \D_2}G_{
\tilde \Delta}^{ \Delta_i}(x_i)
\ee
where the prefactor is,
\be \label{Kdef}
K_{\t \D}^{\D_3, \D_4}= \frac{\pi^{\frac{d}{2}}\G(\frac{d}{2} - \D)\G(\frac{\D - \D_{34}}{2}) \G(\frac{\D+\D_{34}}{2})}{\G(\D)
\G(\frac{\t \D - \D_{34}}{2}) \G(\frac{\t \D +\D_{34}}{2})}~,
\ee
and the conformal block is \cite{Dolan:2000ut}, 
\be \label{Gstandard}
G_{\Delta}^{\Delta_i}(x_i)=\frac{1}{(x_{12}^2)^{\frac{\Delta_1 + \Delta_2}{2}} (x_{34}^2)^{\frac{\Delta_3 + \Delta_4}{2}}}\( \frac{x_{24}^2}{x_{14}^2}\)^{\frac{\Delta_{12}}{2}} \(\frac{x_{14}^2}{x_{13}^2}\)^{\frac{\Delta_{34}}{2}}\, {g}_{\Delta}^{\Delta_i}(u,v)~,
\ee
where the bare conformal block is, 
\be
 g_{\Delta}^{\Delta_i}(u,v) = u^{\frac{\D}{2}}\sum_{n, m=0}^{\infty} \frac{ u^m}{m!} \frac{(1-v)^n}{n!} \frac{(\frac{\D + \D_{34}}{2})_{n+m} (\frac{\D - \D_{12}}{2})_{n+m} (\frac{\D + \D_{12}}{2})_{m} (\frac{\D - \D_{34}}{2})_m}{(\D + 1 - \frac{d}{2})_m (\D)_{2m +n}}~,
\ee
where $u$ and $v$ are the two conformal cross ratios, 
 \be \label{4ptR}
 u= \frac{x_{12}^2 x_{34}^2}{x_{13}^2 x_{24}^2}~, \ \ \ \ \ \ \  v = \frac{x_{14}^2 x_{23}^2}{x_{13}^2 x_{24}^2}~.
 \ee
One can write these cross-ratios as $u = \chi \bar{\chi}$ and $v = (1- \chi) (1-\bar{\chi})$. In $d=2$, these are the $\chi$ and $\bar{\chi}$ that we used in the previous section. In $d=1$, there is only one independent cross ratio, and $\chi = \bar{\chi}$.

\subsection{Five-point block, scalar exchange} \label{5ptd} 
\begin{figure}[t]
\centering
\includegraphics[width=3in]{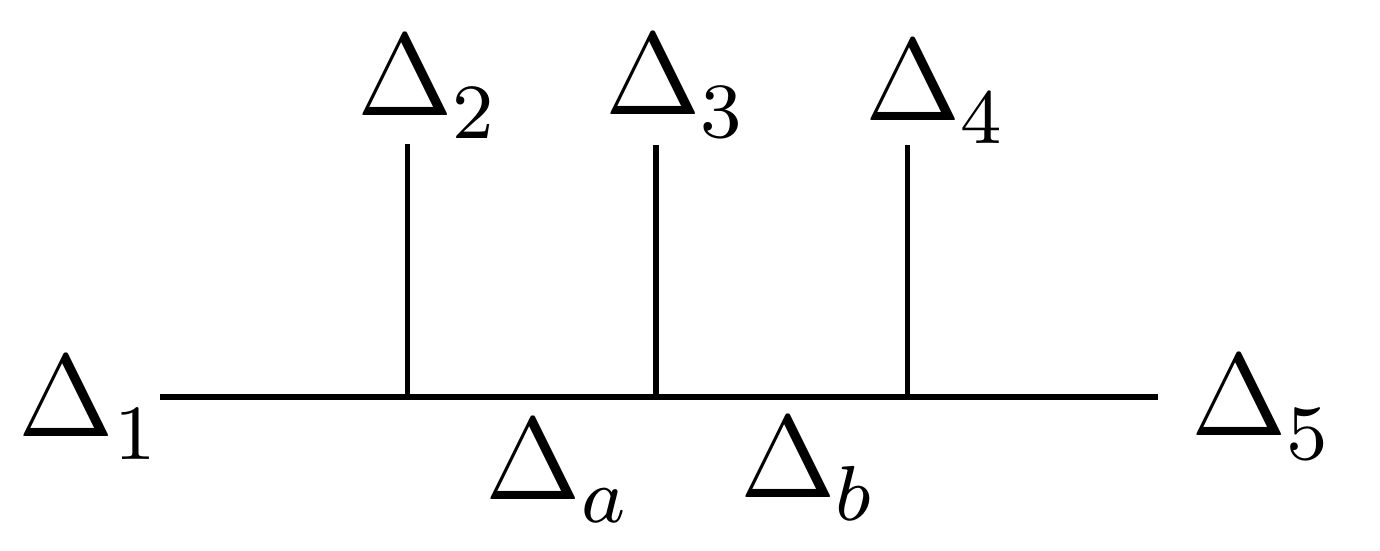}
\caption{The $5$-point partial wave, defined in Eq.~\ref{5ptW}.} \label{5ptdf}
\end{figure}

In this section we compute the five-point conformal  block, with scalar exchange. As in the other cases, we do this by  evaluating the integral expression for the conformal partial wave, and picking out the conformal block in it. 

The five-point conformal partial wave is,
\be \label{5ptW}
\Psi^{\Delta_i}_{\Delta_a,\, \D_b} (x_i) = \int d^d x_a d^d x_b\, \la \mO_1 \mO_2 \mO_a\ra \la \tilde \mO_a \mO_3 \mO_b\ra \la \tilde \mO_b \mO_4 \mO_5\ra~,
\ee
 where $\Delta_i=\Delta_1, \ldots, \Delta_5$ are the dimensions of the external operators  and $\Delta_a$ and $\Delta_b$ are the dimensions of the exchanged operators, $\mO_a$ and $\mO_b$, see Fig.~\ref{5ptdf}. All operators are taken to be scalars. 
 The five-point partial wave, with the integrand written explicitly, is, 
 \be
 \Psi^{\Delta_i}_{\Delta_a, \Delta_b}(x_i) = \int d^d x_a d^d x_b\frac{X_{12}^{\frac{\D_a- \D_1 -\D_2}{2}}X_{45}^{\frac{\tilde \D_b - \D_4 - \D_5}{2}}}{X_{a1}^{\frac{\D_a + \D_{12}}{2}}X_{2a}^{\frac{\D_a - \D_{12}}{2}}X_{3a}^{\frac{\tilde \D_a - \D_b + \D_3}{2}}}\frac{1}{X_{ab}^{\frac{\D_b + \tilde \D_a-\D_3}{2}}}\frac{1}{X_{b3}^{\frac{-\tilde\D_a + \D_b+\D_3}{2}}X_{b4}^{\frac{\tilde \D_b + \D_{45}}{2}} X_{b5}^{\frac{\tilde \D_b - \D_{45}}{2}} }~.
\ee
We first evaluate the 4-point integral over $X_b$, representing it as a double Mellin-Barnes integral, discussed in Appendix.~\ref{MBapp}, Eq.~\ref{I4app}. This gives, 
\bml \label{2025}
 \Psi^{\Delta_i}_{\Delta_a, \Delta_b}(x_i) = \frac{1}{X_{12}^{\frac{\D_1 + \D_2 -\D_a}{2}} X_{35}^{\frac{\D_a - \tilde \D_b + \D_3 }{2}}X_{45}^{\frac{\D_4 + \D_5 - \D_b}{2}} } \frac{\pi^{\frac{d}{2}}}{\G(\frac{\D_b+ \tilde \D_a - \D_3}{2}) \G(\frac{\D_a -\tilde \D_{b}+\D_3}{2})\G(\frac{\t \D_b + \D_{45}}{2}) \G(\frac{\t \D_b -\D_{45}}{2}) }  \\\int \frac{d s}{2\pi i} \frac{d t}{2\pi i}\G(-s)\G(-t) 
\G(\frac{\D_b +\D_{45}}{2} + s+t) \G(\frac{d}{2} - \D_b - s)\G(s+t+ \frac{\D_b - \tilde\D_{a}+\D_3}{2})\G(\frac{\tilde \D_{a}-\D_3 - \D_{45}}{2}-t)\,\\
  \( \frac{X_{45}}{X_{35}}\)^s\(\frac{X_{34}}{X_{35}}\)^t 
\int d^d x_a \frac{1}{X_{1a}^{\frac{\D_a + \D_{12}}{2}} X_{2 a}^{\frac{ \D_a - \D_{12}}{2}} X_{3 a}^{\frac{\tilde \D_a - \D_b  + \D_3}{2}-s} X_{a4}^{\frac{\D_b + \D_{45}}{2}+s+t} X_{a5}^{\frac{\tilde \D_{a}-\D_3 - \D_{45}}{2}-t} }~.
\end{multline}
We now evaluate the integral over $x_a$, representing it as a five-fold Mellin-Barnes integral,  using the result in Appendix.~\ref{MBapp}, Eq.~\ref{5ptD}. Combined with the $s$ and $t$ integrals, this gives us  a seven-fold Mellin-Barnes integral,
\be
 \Psi^{\Delta_i}_{\Delta_a, \Delta_b}(x_i)  = \mathcal{P}^{\D_i}_{\D_a, \D_b} \mathcal{L}^{\D_i}_{\D_a, \D_b}(x_i) \mM^{\D_i}_{\D_a, \D_b}(u_1, v_1, u_2, v_2, w)~,
 \ee
 where $\mathcal{P}^{\D_i}_{\D_a, \D_b}$ is a prefactor of constants,
\be
\mathcal{P}^{\D_i}_{\D_a, \D_b}  = \frac{\pi^d}{\G(\frac{\D_a + \D_{12}}{2}) \G(\frac{\D_a -\D_{12}}{2}) \G(\frac{\tilde \D_b + \D_{45}}{2}) \G(\frac{\tilde \D_b - \D_{45}}{2})} \frac{1}{\G(\frac{\tilde \D_a + \D_b - \D_3}{2})\G(\frac{\D_a - \tilde \D_b + \D_3}{2})}~,
\ee
and $\mathcal{L}^{\D_i}_{\D_a, \D_b}(x_i)$ is the leg factor, 
\be \label{5ptdLeg}
\mathcal{L}^{\D_i}_{\D_a, \D_b}(x_i) =  \frac{1}{(x_{12}^2)^{\frac{\D_1 + \D_2 }{2}} (x_{34}^2)^{\frac{\D_3}{2}} (x_{45}^2)^{\frac{\D_4 + \D_5}{2}}}\(\frac{x_{23}^2}{x_{13}^2}\)^{\frac{\D_{12}}{2}}  \(\frac{x_{24}^2}{x_{23}^2}\)^{\frac{\D_3}{2}} \(\frac{x_{35}^2}{x_{34}^2}\)^{\frac{\D_{45}}{2}}~, 
\ee
and $\mM^{\D_i}_{\D_a, \D_b}$ is the seven-fold Mellin-Barnes integral, 
\bml \label{Miab}
\mM^{\D_i}_{\D_a, \D_b}(u_1, v_1, u_2, v_2, w) = u_1^{\frac{\D_a}{2}}u_2^{\frac{\D_b}{2}}
 \int \frac{d s}{2\pi i} \frac{d t}{2\pi i}\frac{\G(-s)\G(-t) 
 \G(\frac{d}{2} - \D_b - s)\G(s+t+ \frac{\D_b - \tilde\D_{a}+\D_3}{2})}{\G(\frac{\tilde \D_a - \D_b + \D_3}{2}-s)}\\
\int \prod_{i=1}^5 \frac{d s_i}{2\pi i} \G(-s_i)\, \,u_1^{s_1} (v_1-1)^{s_2} (w-1)^{s_3} (v_2-1)^{s_4} u_2^{s_5+s}\, \,  \G(\frac{\D_a {+} \D_b {-} \D_3}{2} +s + \sum_{i=1}^5 s_i)\\
\frac{\G(\frac{\D_a - \D_b + \D_3}{2} -s + s_1 - s_5) \G(\frac{\D_a + \D_{12}}{2} + s_1 + s_2 + s_3) \G(\frac{\D_a - \D_{12}}{2} + s_1 + s_4) }{\G(\D_a + 2s_1 + s_2 + s_3 + s_4)}\\
\frac{\G(\frac{d- 2 \D_a}{2} - s_1 + s_5) \G(\frac{\tilde \D_a - \D_3 - \D_{45}}{2} -t + s_3 + s_4 + s_5)\G(\frac{\D_b + \D_{45}}{2} + s+ t+ s_2 + s_5)}{\G(\frac{\tilde \D_a + \D_b - \D_3}{2} + s + s_2 +s_3+s_4+2s_5)}~,
\end{multline}
 where the five conformal cross-ratios are, 
\be
u_1= \frac{x_{12}^2 x_{34}^2}{x_{13}^2 x_{24}^2}~, \ \ \ v_1 = \frac{x_{14}^2 x_{23}^2}{x_{13}^2 x_{24}^2}~, \ \ \ \ u_2 = \frac{x_{23}^2 x_{45}^2}{x_{24}^2 x_{35}^2}~, \ \ \ \ v_2 = \frac{x_{25}^2 x_{34}^2}{x_{24}^2 x_{35}^2}~, \ \ \ \ w= \frac{x_{15}^2 x_{23}^2 x_{34}^2}{x_{24}^2 x_{13}^2 x_{35}^2}~.
\ee

We are able to evaluate the $t$ integral in (\ref{Miab}), using the first Barnes lemma, leaving us with a six-fold Mellin-Barnes integral. We then close the contours to pick up the poles of the gamma functions, and get that the five-point conformal block is (see Appendix.~\ref{MBapp} for details), 
\be
G_{\D_a, \D_b}^{\D_i}(x_i) =\mathcal{L}^{\D_i}_{\D_a, \D_b}(x_i)\,\, g_{\D_a, \D_b}^{\D_i}(u_1, v_1, u_2, v_2, w)~,
\ee
where the leg factor was given in (\ref{5ptdLeg}), and the bare conformal block is, 
 \bml
g_{\D_a, \D_b}^{\D_i}(u_1, v_1, u_2, v_2, w) =  u_1^{\frac{\D_a}{2}}u_2^{\frac{\D_b}{2}}\sum_{n_i=0}^{\infty}\,\frac{u_1^{n_1}}{n_1!} \frac{(1-v_1)^{n_2} }{n_2!}\frac{(1-w)^{n_3} }{n_3!}\frac{(1-v_2)^{n_4} }{n_4!}\frac{u_2^{n_5}}{n_5!}\,\,(\frac{\D_a {+}\D_b {-} \D_3}{2})_{\sum_{i=1}^5 n_i}\\
\frac{( \frac{\D_a + \D_{12}}{2} )_{ n_1 + n_2 + n_3} (\frac{\D_a - \D_{12}}{2})_{  n_1 + n_4} }{(\D_a )_{2 n_1 + n_2 + n_3 + n_4}}
\frac{(\frac{\D_b - \D_{45}}{2})_{ n_3+n_4+n_5} (\frac{\D_b + \D_{45}}{2} )_{   n_2 + n_5}}{(\D_b )_{  2 n_5 + n_2 + n_3 +n_4}}\frac{(\frac{\D_a - \D_b + \D_3}{2})_{ n_1}(\frac{\D_b - \D_a + \D_3}{2} )_{n_5}}{(1 - \frac{d}{2} + \D_a)_{ n_1} (1 - \frac{d}{2} + \D_b )_{n_5}}\, \\
{}_3 F_2\[\genfrac..{0pt}{}{-n_1,\,\,\,\,\, -n_5,\,\, \,\,\, \frac{ 2- d + \D_a + \D_b - \D_3}{2}}{\frac{\D_a - \D_b - \D_3}{2} {+} 1 {-} n_5,\, \frac{\D_b - \D_a - \D_3}{2} {+} 1 {-}n_1}; 1\]~. 
\end{multline}

We have written the 5-point block in a form that makes the left-right symmetry  of Fig.~\ref{5ptdf} manifest: one can exchange $(1, 2, 3,4, 5) \leftrightarrow (5,4,3,2,1)$,\, \,  $ (\D_a, \D_b)\leftrightarrow (\D_b, \D_a)$.
The form of the block is, as would be expected, more complicated than the 4-point block. It is also qualitatively different, in that the coefficients are not just Pochhammer symbol, but also involve a ${}_3 F_2$. We can not exclude the possibility that there is a different choice of cross-ratios that gives a simpler answer than the one we found. 

For the $4$-point block, there were two independent conformal cross-ratio in any $d\geq 2$. However, for the $5$-point block, there are $4$ independent conformal cross-ratios in $d=2$, but $5$ in $d\geq3$: the cross-ratio we called $w$ is the new one in $d\geq3$.

\section{Discussion} \label{dis}

We  computed the $n$-point conformal blocks, for arbitrary $n$,  in $d=1, 2$,  in the comb channel. The comb channel corresponds to a particular order in which one does the OPE,  and for  $n\geq 6$, there are other channels (which are not related by symmetry). For instance, for $n = 6$,  a different channel is one in which one does the OPE between the first and the second operator, the third and the fourth, and the fifth and the sixth. It is possible to compute the conformal blocks in these other channels, using the same methods  applied here. However, it is not obvious that the answer will take as simple  a form as the one for the comb channel. 

We derived the $5$-point conformal block in arbitrary dimensions, for external and exchanged operators that are scalars. In $d=4$ and $d=6$, for external scalars, the $4$-point conformal block is known to simplify, taking a  similar form as the $d=2$ block \cite{Dolan:2000ut, Dolan:2003hv}. It is conceivable that the $5$-point block could also be simplified further in these dimensions; however, the degree of simplification is limited, since the number of conformal cross-ratios built out of $5$ points is $4$ in $d=2$ but $5$ in $d=4$. For the $4$-point blocks, obtaining the blocks with spinning operators  is nontrivial. One generally uses recursion relations to relate spinning blocks to scalar blocks \cite{Dolan:2003hv, Dolan:2011dv, Costa:2011mg, Costa:2011dw,  Hogervorst:2013sma, Kos:2013tga,   Iliesiu:2015qra, Iliesiu:2015akf, Costa:2016xah,Costa:2016hju, Schomerus:2016epl, Cuomo:2017wme,Echeverri:2015rwa, Penedones:2015aga, Echeverri:2016dun, Karateev:2017jgd, Kravchuk:2017dzd, Lauria:2018klo,Bhatta:2018gjb, Costa:2018mcg}.  Weight-shifting operators \cite{Karateev:2017jgd} can be used to get blocks with either external or internal operators with spin from those with scalars. One could apply weight-shifting operators to the $5$-point block we found. Finally, although we only computed the $5$-point block in $d$ dimensions, one could use the same Mellin-Barnes technique to compute $n$-point blocks;  the result will get progressively more complicated with increasing $n$. It is conceivable, however, that, at least for the comb channel,  there is a pattern which would allow one to eventually guess the $n$-point answer, much as we were able to do in $d=1, 2$.

Our study of $n$-point conformal  blocks was partly motivated by the SYK model \cite{SY, Kitaev,  Rosenhaus:2018dtp, PR, MS, Klebanov:2016xxf, Murugan:2017eto, Kitaev:2017awl, Klebanov:2018fzb}. The $n$-point conformal blocks enter into the solution of SYK in an essential way \cite{GR4}. Harmonic analysis on the conformal group enters as well, and partly due to SYK has undergone a recent revival  \cite{Dobrev:1977qv, Karateev:2018oml, Caron-Huot:2017vep, Simmons-Duffin:2017nub, Liu:2018jhs, Kitaev:2017hnr, Gadde:2017sjg}. It may be useful to extend harmonic analysis on the conformal group to incorporate $n$-point conformal blocks.

\begin{figure}[t]
\centering
\includegraphics[width=3in]{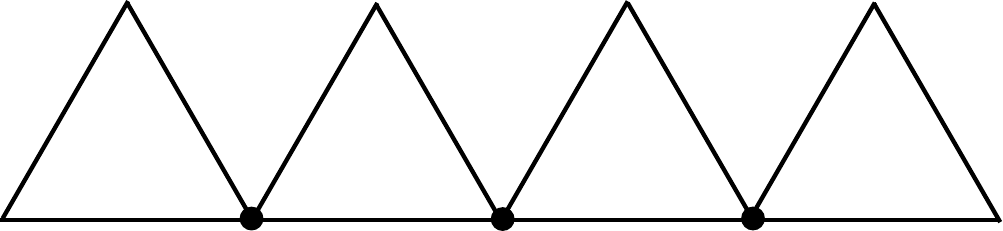}
\caption{The six-point conformal partial wave in the comb channel, represented as a Feynman diagram. Each line is a particular propagator, $1/x^2$ to some power involving the operator dimensions, as specified by the definition of the partial wave, and the three internal points are integrated over. There are six external points which are held fixed. The $n$-point partial wave will look like this figure, but with $n{-}2$ triangles.  }\label{FeynmanDis}
\end{figure}

We  computed the conformal blocks by computing the conformal partial waves, using the shadow formalism. This method turns a conformal block computation into a multi-loop integral. For instance, evaluating the partial waves in the comb channel is equivalent to evaluating the Feynman diagrams shown in Fig.~\ref{FeynmanDis}. There have been many studies of multi-loop Feynman integrals. An interesting and tractable class are those with iterative structure, such as ladder diagrams \cite{Usyukina:1992jd, Usyukina:1993ch}. One can view our result as a computation of a new class of Feynman diagrams. A number of methods have recently been developed for the evaluation of multi-loop Feynman integrals, for example \cite{Isaev:2003tk, Drummond:2006rz, Drummond:2012bg, Basso:2017jwq, Bern:2017gdk , Arkani-Hamed:2017ahv, Bourjaily:2018ycu, Caron-Huot:2018dsv}. Many of these focus on propagators that are $1/x^2$. It would be useful to try to extend these methods to cases in which the propagators contain $1/x^2$ to a non-integer power; this is what occurs in the computation of conformal partial waves.

It is sometimes useful to translate CFT results into AdS statements. Recently, the AdS duals of $4$-point conformal blocks were found to be geodesic Witten diagrams \cite{Hijano:2015zsa}. It would be interesting to find the AdS duals of $n$-point conformal blocks; the explicit form for the blocks that we gave in $d=1, 2$ should make this tractable.

The extensive recent studies of  $4$-point conformal blocks have in part been motivated by their application to  the conformal bootstrap program \cite{Rattazzi:2008pe, Heemskerk:2009pn, Poland:2011ey, ElShowk:2012ht,Komargodski:2012ek,Fitzpatrick:2012yx, Beem:2013qxa, El-Showk:2014dwa, Hartman:2015lfa, Alday:2015ewa, Rychkov:2016iqz, Simmons-Duffin:2016gjk, Poland:2018epd}. It may be useful to study the bootstrap using $n$-point blocks. Although  crossing relations for the four-point function are sufficient for the bootstrap if one incorporates all operators, including  spinning operators,  it may be that if one uses $n$-point functions, then external scalars are sufficient.~\footnote{We thank D.~Gross for this suggestion.}

Finally, even though  two-point and three-point functions of all operators contain all  the CFT data, it may happen that  $n$-point correlation functions of simple operators are  more simple and more natural than  3-point correlation functions of complicated operators.

\section*{Acknowledgements} \noindent I thank  S.~Caron-Huot, D.~Gross, D.~Poland,  D.~Simmons-Duffin, and X.~Yin  for helpful discussions.  
This work was supported  by NSF grant PHY-1606531,  by  NSF grant 1125915 while at the KITP, and by NSF grant PHY-1607611 while at the Aspen Center for Physics.

\appendix

\section{Properties of  the Comb Function $F_K$} \label{appFK}

\subsection{Definition}
The standard one variable hypergeometric function ${}_2 F_1(x)$ is defined as the infinite sum,
\be
{}_2 F_1\[\genfrac..{0pt}{}{a_1, a_2}{c};x\] = \sum_{n=0}^{\infty}\frac{(a_1)_n (a_2)_n}{(c)_n}\frac{x^n}{n!}~, \ \ \ \ \ \  \  \ (a)_n \equiv \frac{\G(a+n)}{\G(a)}~.
\ee
It is common to encounter the single variable generalized hypergeometric function, ${}_p F_q(x)$, defined as, 
\be
{}_p F_q\[\genfrac..{0pt}{}{a_1,\ldots, a_p}{c_1, \ldots, c_q}; x\] = \sum_{n=0}^{\infty}\frac{(a_1)_n \cdots (a_p)_n}{(c_1)_n \cdots (c_q)_n}\frac{x^n}{n!}~.
\ee
We will encounter multivariable hypergeometric functions. The number of simple hypergeometric functions of $k$ variables grows rapidly with $k$, and there does not seem to be a general classification of these beyond three variables. However, for any particular multivariable hypergeometric function, it is straightforward to work out its properties, following standard techniques in the literature for the two variable case, see e.g. \cite{Bailey}.

We define the following function of $k$ variables, which we will refer to as the comb function, 
\bml \label{CombDef1}
\FK{5}{3}{a_1\kk, b_1\kk,\!\ldots\kk, b_{k-1}\kk, a_2}{c_1\kk, \!\ldots\kk, c_k}{x_1\kk,\!\!\ldots\kk, \!x_k} \\
= \sum_{n_1, \ldots, n_k=0}^{\infty} \!\!\frac{(a_1)_{n_1} (b_1)_{n_1 + n_2} (b_2) _{n_2+n_3} \cdots (b_{k-1})_{n_{k-1} +n_k} (a_2)_{n_k}}{(c_1)_{n_1}\cdots (c_k)_{n_k}}\frac{x_1^{n_1}}{n_1!}\! \cdots\!\frac{x_k^{n_k}}{n_k!}~.
\end{multline}
In the slightly degenerate case of one variable, $k=1$, it will be convenient to interpret the definition of $F_K$ to be that of the one variable hypergeometric function, 
\be
\FK{2}{1}{a_1\kk, a_2}{c_1}{x_1} \equiv {}_2 F_1\[\genfrac..{0pt}{}{a_1,\, a_2}{c_1};x_1\]~.
\ee
In the case of two variables, the comb function is,
\be \label{FKAppell}
\FK{3}{2}{a_1\kk, b_1\kk, a_2}{c_1\kk,c_2}{x_1\kk,x_2} = \sum_{n_1, n_2=0}^{\infty}\frac{(a_1)_{n_1} (b_1)_{n_1 + n_2} (a_2)_{n_2}}{(c_1)_{n_1}(c_2)_{n_2}}\frac{x_1^{n_1}}{n_1!} \frac{x_2^{n_2}}{n_2!}~,
\ee
which is the Appell function $F_2$.

\subsubsection*{Splitting identities}
From the definition of the comb function, combined with the trivial identity $(a)_{n+m} = (a+n)_m (a)_n$, one can derive ``splitting'' identities, such as,
\bml \label{Id1}
\FK{5}{3}{a_1\kk,\! b_1\kk, \!\ldots\kk,\! b_{k-1}\kk,a_2}{c_1\kk,\! \! \ldots\kk, c_k}{x_1\kk,\!\! \ldots\kk, \!x_k} \\= \sum_{n_k=0}^{\infty}\frac{(b_{k-1})_{n_k}(a_2)_{n_k}}{(c_k)_{n_k}}\, \frac{x_k^{n_k}}{n_k!}\FK{5}{3}{a_1\kk,\! b_1\kk,\!\! \ldots\kk,\! b_{k-2}\kk,b_{k-1}{+}n_k}{c_1\kk,\! \ldots, c_{k-1}}{x_1\kk,\!\!\ldots\kk,\! x_{k-1}}~.
\end{multline}
Here we have split off the last variable. We may also split off the last two variables, 
\bml \label{Id2}
\FK{5}{3}{a_1\kk,\! b_1\kk,\! \!\ldots\kk,\! b_{k-1}\kk,a_2}{c_1\kk,\! \! \ldots\kk, c_k}{x_1\kk,\!\! \ldots\kk, \!x_k}=\sum_{n_{k-1}=0}^{\infty} 
\frac{(b_{k-2})_{n_{k-1}} (b_{k-1})_{n_{k-1}}}{(c_{k-1})_{n_{k-1}}}\frac{x_{k-1}^{{n_{k-1}}}}{{n_{k-1}}!}\\ \FK{5}{3}{a_1\kk,\! b_1\kk, \!\!\ldots\kk,\! b_{k-3}\kk,b_{k-2}{+}{n_{k-1}}}{c_1\kk,\! \! \ldots\kk, c_{k-2}}{x_1\kk,\!\! \ldots\kk, \!x_{k-2}} 
\FK{2}{1}{b_{k-1}{+}{n_{k-1}}\kk, a_2}{c_k}{x_k}
\end{multline}
Since the  conformal blocks are expressed in terms of the comb function (\ref{nptg0}), these two identities give the following two identities  for the blocks, which we use in the main text: from (\ref{Id1}) we get, 
\bml \label{grel}
g_{\h h_1, \ldots, \h h_{n-3}}^{h_1, \ldots, h_{n}}(\chi_1, \ldots, \chi_{n-3}) \\
= \chi_{n-3}^{\h h_{n-3}}
\sum_{m=0}^{\infty} \frac{(\h h_{n-4} + \h h_{n-3} - h_{n-2})_m (\h h_{n-3} + h_{n} - h_{n-1})_m}{(2 \h h_{n-3})_m}
\, \frac{\chi_{n-3}^{m}}{m!}\, g_{\h h_1, \ldots,  \h h_{n-4}}^{h_1, \ldots, h_{n-2}, \h h_{n-3}+m}(\chi_1, \ldots, \chi_{n-4})~,
\end{multline}
and from (\ref{Id2}) we get,
\bml \label{grel2}
g_{\h h_1, \ldots,  \h h_{n-3}}^{h_1, \ldots, h_{n}}(\chi_1, \ldots, \chi_{n-3}) =\sum_{m=0}^{\infty} \frac{(\h h_{n-5} + \h h_{n-4} - h_{n-3})_m (\h h_{n-4} + \h h_{n-3} - h_{n-2})_m}{(2 \h h_{n-4})_m\, m!}\\ \chi_{n-4}^{m+ \h h_{n-4}}  g_{\h h_1, \ldots, \h h_{n-5}}^{h_1, \ldots, h_{n-3}, \h h_{n-4}+m}(\chi_1, \ldots, \chi_{n-5})\, g_{\h h_{n-3}}^{\h h_{n-4} +m, h_{n-2}, h_{n-1}, h_{n}}(\chi_{n-3})~.
\end{multline}

\subsection{Differential equation representation} \label{CombDiff}
Starting with the definition of the $k$ variable comb function as a sum (\ref{CombDef1}), we derive the set of $k$ differential equations that it satisfies. 

We write the comb function as, 
\be \label{FKSeries}
\FK{5}{3}{a_1\kk, b_1\kk,\!\ldots\kk, b_{k-1}\kk, a_2}{c_1\kk, \!\ldots\kk, c_k}{x_1\kk,\!\!\ldots\kk, \!x_k} = \sum_{n_1,\ldots, n_k=0}^{\infty} A_{n_1, \ldots, n_k}\, x_1^{n_1}\cdots x_k^{n_k}~,
\ee
where the coefficients are given in (\ref{CombDef1}). It is simple to find recursion relations for the coefficients. For instance, the recursion relation which iterates $n_1$ is,
\be
A_{n_1 +1, n_2, \ldots, n_k} = A_{n_1, \ldots, n_k} \frac{(a_1+n_1) (b_1 + n_1+n_2)}{(c_1 + n_1)}\frac{1}{(n_1+1)}~.
\ee
We would like to turn this recursion relation into a differential equation for $F_K$. We define the differential operators, 
\be
\nu_i = x_i \frac{\partial}{\partial x_i}~, \ \ \ \ i=1, \ldots, k~.
\ee
The following equation,
\be \label{Diff1}
\[ x_1 (\nu_1 + a_1) (\nu_1 + \nu_2 + b_1) - \nu_1 (\nu_1 + c_1 - 1)\] F_K=0
\ee
is equivalent to the recursion relation, as one can check by plugging in the series expansion of $F_K$ (here we have suppressed the arguments of $F_K$; they are given in Eq.~\ref{FKSeries}). There are in total $k$ independent recursion relations, which give rise to $k$ differential equations. In particular, the relation,
\be
A_{n_1,\ldots, n_{k-1}, n_k+1} = A_{n_1, \ldots n_k}\frac{(b_{k-1} +n_{k-1} + n_k) (a_2 +n_k)}{(c_k+n_k)}\frac{1}{(n_k+1)}~,
\ee
which iterates $n_k$,
gives the equation, 
\be \label{Diff2}
\[x_k( \nu_{k-1} +\nu_k + b_{k-1})(a_2 + \nu_k) - \nu_k (\nu_k+c_k -1)\]F_K=0~,
\ee
while the relation which iterates $n_2$,
\be \label{n2Rec}
A_{n_1, n_2+1, n_3, \ldots, n_k} = A_{n_1, \ldots, n_k} \frac{(b_1+n_1 + n_2)(b_2 + n_2+n_3)}{(c_2+n_2)}\frac{1}{(n_2+1)}~,
\ee
 gives the equation,
\be
\[x_2(\nu_1+ \nu_2 + b_1)(\nu_2 + \nu_3 +b_2) - \nu_2(\nu_2 + c_2-1)\]F_K=0~.
\ee
Analogous recursion  relations to (\ref{n2Rec}), but involving  the iteration of $n_i$ with $2\leq i\leq k{-}1$ give the equations,
\be \label{Diffk}
\[ x_{a+1}(\nu_a+ \nu_{a+1}+b_a)(\nu_{a+1}+ \nu_{a+2} + b_{a+1}) - \nu_{a+1}(\nu_{a+1} + c_{a+1} -1)\] F_K=0~, \ \ \ \ a = 1, \ldots, k{-}2~.
\ee
In total we have a set of $k$ partial differential  equations, (\ref{Diff1}, \ref{Diff2}, \ref{Diffk}). In the special case that $k=2$, we only have (\ref{Diff1}, \ref{Diff2}), which, written out explicitly, are, 
\bea \label{AppellDif1}
\Big(x_1(1-x_1) \partial_1^2 - x_1 x_2 \partial_1 \partial_2 + \[c_1 - (a_1 + b_1 + 1) x_1\]\partial_1 -a_1 x_2 \partial_2 - a_1 b_1 \Big)F_K&=& 0~, \\  \label{AppellDif2}
\Big(x_2(1-x_2) \partial_2^2 - x_1 x_2 \partial_1 \partial_2 + \[c_2 - (a_2 + b_1 + 1) x_2\]\partial_2 -a_2 x_1 \partial_1 - a_2 b_1 \Big) F_K&=& 0~,
\eea
which are the correct system of differential equations for the Appell function $F_2$.

\subsection{Integral representation}
In this section, we   derive an integral representation of the comb function, by using its definition as a series. 
We start by observing that the ratio of Pochhammer symbols can be written so as to involve a beta function. The using the integral representation of the beta function, we get,
\be \label{PocR}
\frac{(a)_n}{(c)_n} = \frac{\Gamma(c)}{\Gamma(a) \Gamma(c-a)}\frac{\G(a+n) \G(c-a)}{\G(a+n+c-a)} = \frac{\Gamma(c)}{\Gamma(a) \Gamma(c-a)}  \int_0^1 dt\, t^{a+n-1} (1-t)^{c-a-1}~.
\ee
Let us take the series definition of the one variable comb function, in which case it is the standard hypergeometric function ${}_2 F_1$, and make use  of (\ref{PocR}), and then sum the series, to get,
\bea
\FK{2}{1}{a_1\kk, a_2}{c}{x} &=& \frac{\Gamma(c)}{\Gamma(a_2) \Gamma(c{-}a_2)}  \int_0^1 dt\, t^{a_2-1} (1{-}t)^{c-a_2-1} \sum_{n=0}^{\infty}  \frac{(a_1)_n\, (x t)^n}{n!}\\
& =& \frac{\Gamma(c)}{\Gamma(a_2)\Gamma(c{-}a_2)} \int_0^1 d t\,\frac{ t^{a_2-1} (1{-}t)^{c- a_2- 1} }{ (1{-} t x)^{a_1}}~, \label{2F1Int}
\eea
which of course is the standard integral representation of ${}_2 F_1$. Similarly, with the two variable comb function, we start with the series definition (\ref{FKAppell}), apply (\ref{PocR}) twice, and sum the resulting series, to get, 
\be \label{AppellInt}
\FK{3}{2}{a_1\kk, b_1\kk, a_2}{c_1\kk, c_2}{x_1\kk, x_2} 
=\frac{\Gamma(c_1)\Gamma(c_2)}{\Gamma(a_1) \Gamma(c_1{-} a_1)\Gamma(a_2)\Gamma(c_2 {-} a_2)}\int_0^1 dt\,ds\,  \frac{s^{a_1 - 1}(1{-}s)^{c_1 - a_1 - 1}   t^{a_2 - 1}(1{-}t)^{c_2 - a_2 -1}}{(1{-} s x_1 {-} t x_2)^{b_1}}~,
\ee
which reproduces the integral representation of the Appell function  $F_2$. 
For the three variable comb function, we write the series as, 
\be \label{F43}
\FK{4}{3}{a_1\kk,\! b_1\kk,\! b_2\kk,\! a_2}{c_1\kk,\! c_2\kk,\! c_3}{x_1\kk, x_2\kk, x_3} = \sum_{n_2=0}^{\infty} \frac{(b_1)_{n_2} (b_2)_{n_2}}{(c_2)_{n_2}} \frac{x_2^{n_2}}{n_2!}\, \FK{2}{1}{b_1+n_2, a_1}{c_1}{x_1} \FK{2}{1}{b_2 +n_2, a_2}{c_3}{x_3}~,
\ee
and apply the integral representation to each of the two hypergeometric functions, and sum the series to get, 
\begin{multline}
\FK{4}{3}{a_1\kk,\! b_1\kk,\! b_2\kk,\! a_2}{c_1\kk,\! c_2\kk,\! c_3}{x_1\kk, x_2\kk, x_3} =\frac{\Gamma(c_1) \Gamma(c_3)}{\Gamma(a_1) \Gamma(c_1 {-} a_1)\Gamma(a_2)\Gamma(c_3{-}a_2)}\int_0^1 \int_0^1 dt_1 dt_3\, \frac{t_1^{a_1 -1} (1{-}t_1)^{c_1 - a_1- 1}}{ (1{-}t_1 x_1)^{b_1}}\\
 \frac{t_3^{a_2 -1} (1{-}t_3)^{c_3 - a_2- 1} }{ (1{-}t_3 x_3)^{b_2}}\, \FK{2}{1}{b_1, b_2}{c_2}{\frac{x_2}{(1{-}t_1 x_1)(1{-}t_3 x_3)}}~.
\end{multline}
For the $k$ variable comb function, for $k\geq 4$, we write the series in a way that involves a hypergeometric function of $x_1$, and a hypergeometric function of $x_k$, analogous to what we did in the three variable case (\ref{F43}), insert the integral representation of the hypergeometric function, and sum the series, to get, 
\begin{multline} \nn
\FK{5}{3}{a_1\kk, b_1\kk,\!\ldots\kk, b_{k-1}\kk, a_2}{c_1\kk, \!\ldots\kk, c_k}{x_1\kk,\!\!\ldots\kk, \!x_k}  =\frac{\Gamma(c_1) \Gamma(c_{k})}{\Gamma(a_1) \Gamma(c_1 {-} a_1) \Gamma(a_2) \Gamma(c_{k} {-} a_2)}  \int_0^1 dt_1 dt_{k}\\
\frac{t_1^{a_1 -1} (1-t_1)^{c_1 - a_1- 1} }{(1-t_1 x_1)^{b_1}}\, 
 \frac{t_{k}^{a_2 -1} (1-t_{k})^{c_{k} - a_2- 1}}{  (1-t_{k} x_{k})^{b_{k-1}}} 
\FK{3}{3}{  b_1\kk, \ldots\kk, b_{k-1}}{\, \,\,  c_2\kk,\!\! \ldots\kk,\!\! c_{k-1}}{\frac{x_2}{1-t_1 x_1}\kk, x_3\kk,\!\ldots\kk, x_{k-2}\kk, \frac{x_{k-1}}{1 - t_{k} x_{k}}} ~. 
\end{multline}
If $k=4$, then the comb function on the right only has the first and the last arguments.

\section{Mellin-Barnes Integrals} \label{MBapp}

In the computation of the conformal blocks in $d$ dimensions, we encounter integrals of the form,
\be \label{In}
I_n(x_1, \ldots, x_n) = \int d^d x_0 \prod_{i=1}^n \frac{1}{X_{0i}^{a_i}}~, \ \ \ \ X_{0 i} \equiv (x_0 - x_i)^2~,  \ \ \ \sum_{i=1}^n a_i = d~.
\ee
We will refer to these as $n$-point integrals. For general $n$ and general dimension, the best one can do towards evaluating these integrals is turning them into Mellin-Barnes integrals, as we review in this appendix.

We write  each of the factors $1/X_{0 i}^{ a_i}$ in the integrand as an integral, 
\be \label{GInt}
\frac{1}{X_{0 i}^{a_i}} = \frac{1}{\G(a_i)} \int_0^{\infty} \frac{d \lam_i}{\lam_i} \lam_i^{a_i}\, e^{- \lam_i X_{0i}}~,
\ee
and then evaluate the $x_0$ integral, by completing the square. This gives, 
\be \label{In2}
I_n = \frac{\pi^{\frac{d}{2}}}{\prod_{i=1}^n \G(a_i)}\int_0^{\infty} \prod_{i=1}^n \frac{d\lam_i}{\lam_i} \lam_i^{a_i} \, \, \frac{1}{\Lambda^{\frac{d}{2}}}\, \exp\( - \frac{1}{\Lambda} \sum_{1\leq i<j\leq n} \lam_i \lam_j X_{i j}\)~,
\ee
where $\Lambda = \sum_{i=1}^n \lam_i$. The $n$-point integral is conformally invariant if $\sum_{i=1}^n a_i = d$. In this case, we may replace $\Lambda$ by $\Lambda = \sum_{i=1}^n \alpha_i \lam_i$, provided $\alpha_i>0$ \cite{Symanzik:1972wj}.

Let us complete the evaluation of the integral, in  the case of the $3$-point integral. We choose $\Lambda = \lam_3$. With this choice, the integral over $\lam_3$ that appears  in (\ref{In2}) is,
\be
\int_0^{\infty} \frac{d \lam_3}{\lam_3}\, \lam_3^{a_3 - \frac{d}{2}}\, \exp\( - \frac{\lam_1 \lam_2}{\lam_3} X_{12}\) =  \frac{\G(\frac{d}{2} - a_3)}{(X_{12} \lam_1 \lam_2)^{\frac{d}{2} - a_3}}~,
\ee
where we got the second line by a change of variables $\lam_3 \rightarrow 1/\lam_3$ followed by application of (\ref{GInt}). We then perform the $\lam_1$ and $\lam_2$ integrals using (\ref{GInt}), to find, 
\be \label{ST}
I_3 = \pi^{\frac{d}{2}} \prod_{i=1}^3\frac{\G(\frac{d}{2} - a_i)}{\G(a_i)}\, \frac{1}{X_{12}^{\frac{d}{2} - a_3} X_{13}^{\frac{d}{2} - a_2} X_{23}^{\frac{d}{2} - a_1}} ~.
\ee
This is the famous star-triangle relation. 

For evaluating $I_n$ for $n\geq 4$, we would like to use the same procedure. However, there are now more terms in the exponent. For some of these, we use the following representation of the exponential,
\be \label{eMB}
e^{- x} = \int \frac{ds}{2\pi i}\,  \G(-s )\, x^s~,
\ee
where the contour of integration is along a line parallel to, and to the left of, the imaginary axis: $c{-}i\infty<s<c{+}i \infty$ with $c<0$.

\subsection{$4$-point integral}
We now evaluate the $4$-point integral, (\ref{In2}) with $n=4$.
We let $\Lambda = \lam_4$ and write, 
\be \nonumber
\exp( - \frac{\lam_1 \lam_2}{\lam_4} X_{12}) = \int \frac{d s}{2\pi i} \G(-s)\( \frac{\lam_1 \lam_2}{\lam_4} X_{12}\)^s~,\, \, \, \ \ \ \ \ \ 
\exp( - \frac{\lam_2 \lam_3}{\lam_4} X_{23}) = \int \frac{d t}{2\pi i}\G(-t) \( \frac{\lam_2 \lam_3}{\lam_4} X_{23}\)^t~.
\ee
Hence $I_4$ becomes,
\bea 
 \!\!\!\!\!\!I_4 &=& \frac{\pi^{\frac{d}{2}}}{\prod_{i=1}^4 \G(a_i)} \int \frac{d s}{2\pi i} \frac{d t}{2\pi i}\G(-s)\G(-t)X_{12}^s X_{23}^t\int_0^{\infty}\prod_{i=1}^4 \frac{d\lam_i}{\lam_i} \lam_i^{a_i} \frac{1}{\lam_4^{\frac{d}{2}}}\( \frac{\lam_1 \lam_2}{\lam_4} \)^s\ \( \frac{\lam_2 \lam_3}{\lam_4} \)^t \\ \nn
&&\exp\( -\lam_1 X_{14} - \lam_2 X_{24} - \lam_3 X_{34}\) \exp\(- \frac{\lam_1\lam_3}{\lam_4} X_{13}\)~.
\eea
We successively do   the $\lam_4, \lam_3, \lam_2, \lam_1$ integrals, to obtain, 
\begin{multline} \label{I4app}
I_4 = X_{13}^{a_4 - \frac{d}{2}} X_{34}^{\frac{d}{2} - a_3 - a_4} X_{24}^{-a_2} X_{14}^{\frac{d}{2} - a_1 - a_4}\,  \frac{\pi^{\frac{d}{2}}}{\prod_{i=1}^4 \G(a_i)}  \int \frac{d s}{2\pi i} \frac{d t}{2\pi i}\G(-s)\G(-t) \\ 
\G(\frac{d}{2} - a_4 + s+t) \G(a_3 + a_4 - \frac{d}{2} - s)\G(s+t+a_2)\G(a_1 + a_4 - \frac{d}{2} -t)\, u^s v^t~,
\end{multline}
where $u$ and $v$ are the two conformal cross-ratios, 
\be
u= \frac{X_{12} X_{34}}{X_{13} X_{24}}~, \ \  \ \ \ \ v= \frac{X_{14} X_{23}}{X_{13}X_{24}}~.
\ee
One can see that we made the choice of $\Lambda = \lam_4$ in order to obtain the result in terms of the standard $u$ and $v$ cross-ratios. We have also used  $\sum_{i=1}^4 a_i = d$ to simplify expressions. 

If one closes the contours of the $s$ and $t$ integral, one picks up the poles from the gamma functions, and obtains a double sum. In some contexts, we want to study the OPE regime, in which $u\rightarrow 0$ and $v\rightarrow 1$. For this, it is better to expand in $(1-v)$, rather than $v$.  

To do this we will make use of several identities. First, the Mellin-Barnes expansion, 
\be \label{MBxy}
\frac{1}{(x+y)^a} = \frac{1}{\G(a)}\int \frac{ d s }{2\pi i}\, \G(-s) \G(s+a) x^s y^{- a- s}~.
\ee
If $a>0$, the contour can be chosen to run along the imaginary axis $-i \infty<s<i \infty$. If $a<0$, then some of the poles of $\G(s+a)$ (located at $s= - a-n$) are to the right of the imaginary axis. Thus, if $a\leq0$, one should add a small imaginary piece to $a$, and deform the contour of integration to always be to the right of the poles of $\G(s+a)$ and to the left of the poles of $\G(-s)$. This is the correct contour, since with it one recovers the  Taylor expansion of $(x+y)^{-a}$. Specifically, 
if $x<y$, one can close the contour to the right, picking up the poles of $\G(-s)$,  recovering the Taylor expansion of $y^{-a}(1+\frac{x}{y})^{-a}$. If $x>y$, one can close the contour to the left, picking up the poles of $\G(s+a)$, and recovering the Taylor expansion of $x^{-a}(1+ \frac{y}{x})^{-a}$. The advantage of using the Mellin-Barnes representation over a Taylor expansion is that there is one expression which is valid regardless of if $x$ is less than or greater than $y$. 

Applying the Mellin-Barnes expansion (\ref{MBxy}) to  $v = 1-(1-v)$ gives, 
\be \label{vMB}
v^{t} = \frac{1}{\G( -t )}\int \frac{d \bar t}{2\pi i} \G( -\bar t) \G( \bar t- t) (v - 1)^{\bar t}~.
\ee
We will also make use of the first Barnes lemma, 
\be \label{Barnes}
\int_{- i\infty}^{i\infty} \frac{d s}{2\pi i} \G(a_1+s) \G(a_2+s) \G(b_1-s) \G(b_2-s) = \frac{\G(a_1 + b_1) \G(a_1 + b_2) \G(a_2 + b_1) \G(a_2+b_2)}{\G(a_1 + a_2 + b_1 + b_2)}~.
\ee
We now return to the result (\ref{I4app}) for the 4-point integral,  insert (\ref{vMB}), perform the integral over $t$  by use of the first Barnes lemma, 
and then  relabel $\bar t \rightarrow t$, to get, 
\begin{multline} \label{I4appv2}
I_4 = X_{13}^{a_4 - \frac{d}{2}} X_{34}^{\frac{d}{2} - a_3 - a_4} X_{24}^{-a_2} X_{14}^{\frac{d}{2} - a_1 - a_4}\,  \frac{\pi^{\frac{d}{2}}}{\prod_{i=1}^4 \G(a_i)}  \int \frac{d s}{2\pi i} \frac{d t}{2\pi i}\G(-s)\G(-t) \\ 
\,\frac{ \G(a_3 + a_4 - \frac{d}{2} - s)\G(a_1 +s) \G(\frac{d}{2} - a_4+s+t)\G(\frac{d}{2} - a_3 + s) \G(a_2+s+t)}{\G(2s + t + a_1 + a_2)}\,  u^s (v-1)^t~.
\end{multline}
We may close the contours on the right, to get, 
\begin{multline} \label{I4appv3}
I_4 = X_{13}^{a_4 - \frac{d}{2}} X_{34}^{\frac{d}{2} - a_3 - a_4} X_{24}^{-a_2} X_{14}^{\frac{d}{2} - a_1 - a_4}\,  \frac{\pi^{\frac{d}{2}}}{\prod_{i=1}^4 \G(a_i)}  \sum_{m, n=0}^{\infty} \frac{(-u)^m}{m!} \frac{(1-v)^n}{n!}\\
\Big[\,\frac{ \G(a_3 + a_4 - \frac{d}{2} - m)\G(a_1 +m) \G(\frac{d}{2} - a_4+m+n)\G(\frac{d}{2} - a_3 + m) \G(a_2+m+n)}{\G(2m + n+ a_1 + a_2)}\,  \\
+u^{a_3 + a_4 - \frac{d}{2}}\,\frac{ \G (\frac{d}{2}-a_3 - a_4 -m)\G(\frac{d}{2}-a_2+m) \G(a_3+m+n)\G(a_4+ m) \G(\frac{d}{2}-a_1 + m+n)}{\G(a_3+a_4+ 2m+n)}
\Big]~,
\end{multline}
where the first term came from the poles of $\G(-s)$, located at $s= m$, and the second term came from the poles of $\G(a_3 + a_4 - \frac{d}{2} - s)$,  located at $s= a_3 +a_4 - \frac{d}{2} +m$. Both terms picked up the poles from $\G(-t)$ at $t = n$. We may move terms between the numerator and denominator by using the identity, 
\be
\Gamma(a - m) = (-1)^m \frac{\Gamma(a) \Gamma(1-a)}{\Gamma(1- a+m)}~.
\ee

\subsection{$5$-point integral}
Next, we consider the $5$-point integral. We would like to maintain symmetry between $(1,2)$ and $(4,5)$, so in (\ref{In2}) we take $\Lambda = \lam_3$. We apply the integral representation (\ref{eMB}) to some of the exponentials, to write $I_5$ as,  
\bml
\!\!\!\!\!\! I_5 = \frac{\pi^{\frac{d}{2}}}{\prod_{i=1}^5 \G(a_i)} \int \prod_{i=1}^5 \frac{d s_i}{2\pi i}\, \G(- s_i)\, X_{12}^{s_1} X_{14}^{s_2} X_{15}^{s_3} X_{25}^{s_4}\, X_{45}^{s_5} \int_0^{\infty} \prod_{i=1}^5 \frac{d \lam_i}{\lam_i}\, \lam_i^{a_i} \frac{1}{\lam_3^{\frac{d}{2}}} \(\frac{\lam_1 \lam_2}{\lam_3}\)^{s_1}\! \! \(\frac{\lam_1 \lam_4}{\lam_3}\)^{s_2}\!\! \(\frac{\lam_1 \lam_5}{\lam_3}\)^{s_3}\\ \nn
\(\frac{\lam_2 \lam_5}{\lam_3}\)^{s_4} \(\frac{\lam_4 \lam_5}{\lam_3}\)^{s_5}\, \exp\( - \lam_1 X_{13} - \lam_2 X_{23} - \lam_4 X_{34} - \lam_5 X_{35}\)\, \exp\( - \frac{\lam_2 \lam_4}{\lam_3} X_{24}\)~.
\end{multline}
Notice that we have the symmetry, 
\be
(1, 2, 3,4,5) \leftrightarrow (5,4,3,2,1)~,
\ee
if we act on the $\lambda_i, s_i, a_i$. 
We successively do the $\lam_3, \lam_1, \lam_5, \lam_2, \lam_4$ integrals, to get, 
\bml
\!\!\!\!\!\! I_5 = \frac{1}{X_{24}^{\frac{d}{2} - a_3} X_{13}^{a_1} X_{35}^{a_5} X_{23}^{a_2 + a_3 - \frac{d}{2}} X_{34}^{a_3 + a_4 - \frac{d}{2}}} \frac{\pi^{\frac{d}{2}}}{\prod_{i=1}^5 \G(a_i)} \int \prod_{i=1}^5 \frac{d s_i}{2\pi i}\,\, \G(- s_i)\,\G(\sum_{i=1}^5 s_i + \frac{d}{2} - a_3)
\G(s_1 + s_2 + s_3 + a_1)\\\G(s_3 + s_4 + s_5+a_5)\G(a_2 + a_3 {-} s_2 {-} s_3{ -} s_5{ -}\frac{d}{2})\G(a_4 + a_3 {-} s_1 {-} s_3 {-} s_4 {-} \frac{d}{2})\,\,  u_1^{s_1}\, v_1^{s_2}\,w^{s_3}\,  v_2^{s_4}\, u_2^{s_5}\,~,
\end{multline}
where the five conformal cross-ratios are, 
\be
u_1= \frac{X_{12} X_{34}}{X_{13} X_{24}}~, \ \ \ v_1 = \frac{X_{14} X_{23}}{X_{13} X_{24}}~, \ \ \ \ u_2 = \frac{X_{23} X_{45}}{X_{24}X_{35}}~, \ \ \ \ v_2 = \frac{X_{25} X_{34}}{X_{24}X_{35}}~, \ \ \ \ w= \frac{X_{15} X_{23} X_{34}}{X_{24} X_{13} X_{35}}~.
\ee

Another form of $I_5$ that will be useful  is  obtained by applying (\ref{vMB}) to $ v_1^{s_2}$, to $W^{s_3}$, and to $  v_2^{s_4}$. Performing the integrals over $s_2, s_3, s_4$ and relabeling $\bar{s}_2, \bar{s}_3, \bar{s}_4 \rightarrow s_2, s_3, s_4$, respectively, gives,~\footnote{In more detail, we first do the  $s_2$ integral using the Barnes lemma. Then we change variables $s_3 \rightarrow s_3 - s_4$ and do the $s_4$ integral, and then the $s_3$ integral, using the Barnes lemma.}
\bml \label{5ptD}
I_5 = \frac{1}{X_{24}^{\frac{d}{2} - a_3} X_{13}^{a_1} X_{35}^{a_5} X_{23}^{a_2 + a_3 - \frac{d}{2}} X_{34}^{a_3 + a_4 - \frac{d}{2}}} \frac{\pi^{\frac{d}{2}}}{\prod \G(a_i)} \int \prod \frac{d s_i}{2\pi i}\, \G({-} s_i)\ \G(\sum s_i +\frac{d}{2}  - a_3) \\
\frac{\G(s_1 -s_5 + a_1 + a_2 + a_3 - \frac{d}{2})\G(s_1 + s_2 + s_3 + a_1) \G(s_1 + s_4 + a_2)}{\G(2 s_1 + s_2 +s_3 +s_4 + a_1 + a_2)}\\
\frac{\G(-s_1+s_5+a_3 + a_4+ a_5 - \frac{d}{2})\G(s_3 + s_4 + s_5 + a_5) \G(s_2 + s_5+a_4)}{\G(s_2 + s_3 + s_4 + 2s_5 + a_4 + a_5)}
\\
u_1^{s_1} u_2^{s_5} (v_1 - 1)^{s_2} (w-1)^{s_3} (v_2-1)^{s_4}~.
\end{multline}

\section{Details of the $d$ dimensional Five-Point Block} \label{5dDetails}

In this appendix we fill in the details of the computation  in Sec.~\ref{5ptd} of the five-point conformal block with scalar exchange in $d$ dimensions. 

In Sec.~\ref{5ptd} we computed  the five-point partial wave, expressing it as a seven-fold Mellin-Barnes integral, $\mM^{\D_i}_{\D_a, \D_b}$ given in (\ref{Miab}). We can simplify $\mM^{\D_i}_{\D_a, \D_b}$ by performing the $t$ integral, using the first Barnes lemma (\ref{Barnes}), to get, 
\bml
\mM^{\D_i}_{\D_a, \D_b}(u_1, v_1, u_2, v_2, w) = u_1^{\frac{\D_a}{2}}u_2^{\frac{\D_b}{2}}
 \int \frac{d s}{2\pi i} \frac{\G(-s)
 \G(\frac{d}{2} - \D_b - s)\G(s+ \frac{\D_b - \tilde\D_{a}+\D_3}{2})}{\G(\frac{\tilde \D_a - \D_b + \D_3}{2}-s)}\\
\int \prod_{i=1}^5 \frac{d s_i}{2\pi i} \G(-s_i)\, \,u_1^{s_1} (v_1-1)^{s_2} (w-1)^{s_3} (v_2-1)^{s_4} u_2^{s_5+s}\, \,  \G(\frac{\D_a {+}\D_b {-} \D_3}{2} +s + \sum_{i=1}^5 s_i)\\
\frac{\G(\frac{\D_a - \D_b + \D_3}{2} -s + s_1 - s_5) \G(\frac{\D_a + \D_{12}}{2} + s_1 + s_2 + s_3) \G(\frac{\D_a - \D_{12}}{2} + s_1 + s_4) }{\G(\D_a + 2s_1 + s_2 + s_3 + s_4)}\\
\frac{\G(\frac{\D_b - \D_{45}}{2} + s+s_3+s_4+s_5)\G(\frac{d- 2 \D_a}{2} - s_1 + s_5)\G(\frac{\D_b + \D_{45}}{2} + s+  s_2 + s_5)}{\G(\D_b + 2 s + 2s_5 + s_2 + s_3 +s_4)}~.
\end{multline}
To simplify further, we could change variables $s_5 \rightarrow s_5 -s$, and then do the $s$ integral. The form of the $s$ integral is structurally similar to what appears in the second Barnes lemma, with five gamma functions in the numerator and one gamma function in the denominator; however, the coefficients we have in the denominator are different from those required by the second Barnes lemma, so it is inapplicable. The best we can do is to perform the $s$ integral  by closing the contour and  writing it as a sum of two ${}_3 F_2$'s. This is not really a simplification, so we will not do this. 

We can close all 6 contour integrals,  picking up the poles of the gamma functions, to write the expression as a sum of several six-fold sums. All contours are closed to the right. For the variables $s_2, s_3, s_4$, the only poles come from $\G(-s_2), \G(-s_3), \G(-s_4)$, at $s = n_2$,  $s=n_3$, $s=n_4$, respectively. For the other three variables, $s, s_1, s_5$, there are multiple gamma functions which give poles on the right. 
However, we are only interested in picking out the conformal block. The first term in the conformal block, in powers of $u_1$ and $ u_2$, is $u_1^{\frac{\D_a}{2}} u_2^{\frac{\D_b}{2}}$. Thus, we should look at the poles coming from $\G(-s), \G(-s_1), \G(-s_5)$, which are at $s =n$, $s_1 = n_1$, $s_5= n_5$, respectively. We will denote this piece of $\mM^{\D_i}_{\D_a, \D_b}$ by $M^{\D_i}_{\D_a, \D_b}$. We have,
\bml
M^{\D_i}_{\D_a, \D_b}(u_1, v_1, u_2, v_2, w) = u_1^{\frac{\D_a}{2}}u_2^{\frac{\D_b}{2}}\sum_{n, n_i=0}^{\infty}\,\frac{(-u_1)^{n_1}}{n_1!} \frac{(1-v_1)^{n_2} }{n_2!}\frac{(1-w)^{n_3} }{n_3!}\frac{(1-v_2)^{n_4} }{n_4!}\frac{(-u_2)^{n_5+n}}{n_5!\, n!}\\
\frac{
 \G(\frac{d}{2} - \D_b - n)\G(n+ \frac{\D_b - \tilde\D_{a}+\D_3}{2}) \G(\frac{\D_a {+}\D_b {-} \D_3}{2} +n+ \sum_{i=1}^5 n_i)}{\G(\frac{\tilde \D_a - \D_b + \D_3}{2}-n)}
\, \, \\
\frac{\G(\frac{\D_a - \D_b + \D_3}{2} -n + n_1 - n_5) \G(\frac{\D_a + \D_{12}}{2} + n_1 + n_2 + n_3) \G(\frac{\D_a - \D_{12}}{2} + n_1 + n_4) }{\G(\D_a + 2 n_1 + n_2 + n_3 + n_4)}\\
\frac{\G(\frac{\D_b - \D_{45}}{2} + n+n_3+n_4+n_5)\G(\frac{d- 2 \D_a}{2} - n_1 + n_5)\G(\frac{\D_b + \D_{45}}{2} + n+  n_2 + n_5)}{\G(\D_b + 2 n + 2 n_5 + n_2 + n_3 +n_4)}~.
\end{multline}
We change variables $n_5\rightarrow n_5-n$ to write this as,
\bml \label{MDiab}
M^{\D_i}_{\D_a, \D_b}= u_1^{\frac{\D_a}{2}}u_2^{\frac{\D_b}{2}}\sum_{ n_i=0}^{\infty}\,\frac{(-u_1)^{n_1}}{n_1!} \frac{(1-v_1)^{n_2} }{n_2!}\frac{(1-w)^{n_3} }{n_3!}\frac{(1-v_2)^{n_4} }{n_4!}\frac{(-u_2)^{n_5}}{n_5!}
\ \G(\frac{\D_a {+}\D_b {-} \D_3}{2} + \sum_{i=1}^5 n_i)
\, \, \\
\frac{ \G(\frac{\D_a + \D_{12}}{2} + n_1 + n_2 + n_3) \G(\frac{\D_a - \D_{12}}{2} + n_1 + n_4) }{\G(\D_a + 2 n_1 + n_2 + n_3 + n_4)}
\frac{\G(\frac{\D_b - \D_{45}}{2} +n_3+n_4+n_5)\G(\frac{\D_b + \D_{45}}{2} +   n_2 + n_5)}{\G(\D_b +  2 n_5 + n_2 + n_3 +n_4)}\mathcal{S}~,
\end{multline}
where, 
\bml
\mathcal{S} = \G(\frac{\D_a - \D_b + \D_3}{2} + n_1 - n_5)\\
\sum_{n=0}^{n_5}\frac{n_5!}{(n_5-n)!n!}\frac{ \G(n+ \frac{\D_b - \tilde\D_{a}+\D_3}{2})\G(\frac{d}{2} - \D_b - n)\G(\frac{d- 2 \D_a}{2} - n_1 + n_5-n)}{\G(\frac{\tilde \D_a - \D_b + \D_3}{2}-n)}~.
\end{multline}
We can write $\mathcal{S}$  in terms of the hypergeometric function ${}_3 F_2$, 
\bml
\mathcal{S} = \frac{\G(\frac{\D_a - \D_b + \D_3}{2} {+} n_1 {-} n_5)\G(\frac{d}{2} {-} \D_a {-} n_1 +n_5)\G(\frac{d}{2} {-} \D_b) \G(\frac{\D_ a+ \D_b + \D_3 - d}{2})}{\G(\frac{d - \D_a - \D_b + \D_3}{2})}\, \\
{}_3 F_2\[\genfrac..{0pt}{}{-n_5,\,\, \,  \frac{ 2- d - \D_3 + \D_a + \D_b}{2},\,\, \, \frac{\D_a + \D_b + \D_3 -d}{2}}{1{+} \D_a {-} \frac{d}{2} {+} n_1 {-} n_5,\,  1 {+} \D_b {-} \frac{d}{2}}; 1\]~.
\end{multline}
We would like to write $\mathcal{S}$ in a more symmetric form. For this we make use of the following identity for ${}_3F_2$  (see Eq. 2.5.11 of \cite{Slater}),
\be
{}_3 F_2 \[\genfrac..{0pt}{}{a, b, -n}{e,  f}; 1\] = \frac{(e-a)_n (f-a)_n}{(e)_n (f)_n}\, \,  {}_3 F_2 \[\genfrac..{0pt}{}{1{-}s,\,\,\, a,\,\,\, {-}n}{1{+}a{-} e{-}n, 1{+}e{-}f{-}n}; 1\]~, \ \ \ \ s = e{+}f{ -} a{-} b{+}n~,
\ee
enabling us to write $\mathcal{S}$ as,
  \bml
 \mathcal{S} =(-1)^{n_1 +n_5}\frac{\G(\frac{\D_a - \D_b + \D_3}{2})\G(\frac{d}{2} - \D_a)\G(\frac{d}{2}-\D_b)\G(\frac{\D_a + \D_b + \D_3 - d}{2})}{\G(\frac{- \D_a - \D_b + \D_3 +d}{2})}\frac{(\frac{\D_a - \D_b + \D_3}{2})_{ n_1}(\frac{\D_b - \D_a + \D_3}{2} )_{n_5}}{(1 - \frac{d}{2} + \D_a)_{ n_1} (1 - \frac{d}{2} + \D_b )_{n_5}}\, \\
 {}_3 F_2\[\genfrac..{0pt}{}{-n_1,\,\,\,\,\, -n_5,\,\, \,\,\, \frac{ 2- d + \D_a + \D_b - \D_3}{2}}{\frac{\D_a - \D_b - \D_3}{2} {+} 1 {-} n_5,\, \frac{\D_b - \D_a - \D_3}{2} {+} 1 {-}n_1}; 1\]~.
 \end{multline}
We insert this form of $\mathcal{S}$ into the expression for $M^{\D_i}_{\D_a, \D_b}$ to get,
\be \label{Mmm}
M^{\D_i}_{\D_a, \D_b}(u_1, v_1, u_2, v_2, w) =  K_{\tilde \D_a}^{\D_3, \D_b} K_{\tilde \D_b}^{\D_4, \D_5}\,\,  g_{\D_a, \D_b}^{\D_i}(u_1, v_1, u_2, v_2, w)~,
\ee
where the factor $K_{\tilde \D_a}^{\D_3, \D_b}$ is a ratio of  gamma functions, defined in (\ref{Kdef}), and $g_{\D_a, \D_b}^{\D_i}$ is, 
\bml
g_{\D_a, \D_b}^{\D_i}(u_1, v_1, u_2, v_2, w) =  u_1^{\frac{\D_a}{2}}u_2^{\frac{\D_b}{2}}\sum_{n_i=0}^{\infty}\,\frac{u_1^{n_1}}{n_1!} \frac{(1-v_1)^{n_2} }{n_2!}\frac{(1-w)^{n_3} }{n_3!}\frac{(1-v_2)^{n_4} }{n_4!}\frac{u_2^{n_5}}{n_5!}\,\,(\frac{\D_a {+}\D_b {-} \D_3}{2})_{\sum_{i=1}^5 n_i}\\
\frac{( \frac{\D_a + \D_{12}}{2} )_{ n_1 + n_2 + n_3} (\frac{\D_a - \D_{12}}{2})_{  n_1 + n_4} }{(\D_a )_{2 n_1 + n_2 + n_3 + n_4}}
\frac{(\frac{\D_b - \D_{45}}{2})_{ n_3+n_4+n_5} (\frac{\D_b + \D_{45}}{2} )_{   n_2 + n_5}}{(\D_b )_{  2 n_5 + n_2 + n_3 +n_4}}\frac{(\frac{\D_a - \D_b + \D_3}{2})_{ n_1}(\frac{\D_b - \D_a + \D_3}{2} )_{n_5}}{(1 - \frac{d}{2} + \D_a)_{ n_1} (1 - \frac{d}{2} + \D_b )_{n_5}}\, \\
{}_3 F_2\[\genfrac..{0pt}{}{-n_1,\,\,\,\,\, -n_5,\,\, \,\,\, \frac{ 2- d + \D_a + \D_b - \D_3}{2}}{\frac{\D_a - \D_b - \D_3}{2} {+} 1 {-} n_5,\, \frac{\D_b - \D_a - \D_3}{2} {+} 1 {-}n_1}; 1\]~.
\end{multline}
We identify $g_{\D_a, \D_b}^{\D_i}$ as the bare five-point conformal block. In addition, the coefficient in (\ref{Mmm}) is the correct coefficient, as one can establish by evaluating the partial wave in the OPE limit in which the integrals reduce to star-triangle integrals (\ref{ST}).

\bibliographystyle{utphys}

\end{document}